\documentclass[sigconf]{acmart}
\newcommand{\EuclideanOne}{\mathbb{R}}
\newcommand{\EuclideanOnePos}{\mathbb{R^+}}
\newcommand{\EuclideanTwo}{\mathbb{R}^{2}}
\newcommand{\EuclideanThree}{\EuclideanTwo}
\newcommand{\TimeStamp}{t}

\newcommand{\RepTimeStamp}{\bar{t}}
\newcommand{\TimeStampStart}{t_s}
\newcommand{\TimeStampEnd}{t_e}
\newcommand{\TimeInterval}{T}
\newcommand{\Rotation}{\mathbf{r}}
\newcommand{\RotationStart}{\mathbf{\Rotation_s}}
\newcommand{\RotationEnd}{\mathbf{\Rotation_e}}

\newcommand{\AngularVelocity}{\boldsymbol{\omega}}
\newcommand{\AngularVelocityDeriv}{\dot{\AngularVelocity}}
\newcommand{\AngularAcceleration}{\boldsymbol{\alpha}}
\newcommand{\AngularAccelerationPitch}{\AngularAcceleration_\PitchStart}
\newcommand{\AngularAccelerationYaw}{\AngularAcceleration_\YawStart}

\newcommand{\PassiveTorque}{\mathbf{\mathcal{T}_{p}}}
\newcommand{\ActiveTorque}{\mathbf{\mathcal{T}_{a}}}

\newcommand{\Inertia}{I}

\newcommand{\TorqueToEnergy}{\mathcal{E}}

\newcommand{\CombinedEnergyFunctional}{\mathcal{H}_m}
\newcommand{\CombinedEnergy}{\mathrm{H}_c}

\newcommand{\CoverageEnergy}{\mathrm{C}}

\newcommand{\PilotScanPath}{\mathcal{S}}
\newcommand{\PilotTargetPos}{\mathbf{r}}
\newcommand{\PilotConditionSet}{\mathcal{R}}

\newcommand{\PilotConditionSize}{N_\PilotConditionSet}

\newcommand{\PilotTargetPosStart}{\mathbf{r}}
\newcommand{\PilotTargetPosDelta}{\mathbf{\Delta\PilotTargetPosStart}}

\newcommand{\Pitch}{p}
\newcommand{\Yaw}{y}

\newcommand{\PitchStart}{\Pitch}
\newcommand{\YawStart}{\Yaw}

\newcommand{\PitchDelta}{{\Delta\Pitch}}
\newcommand{\YawDelta}{{\Delta\Yaw}}

\newcommand{\conditionRand}{\mathbf{RND}}
\newcommand{\conditionMin}{\mathbf{MIN}}
\newcommand{\conditionMax}{\mathbf{MAX}}

\newcommand{\conditionMaxVMin}{\mathbf{C_1}}
\newcommand{\conditionMaxVRand}{\mathbf{C_2}}
\newcommand{\conditionMinVRand}{\mathbf{C_3}}

\newcommand{\expenditureShort}{MCL\xspace}
\newcommand{\activationShort}{MCL\xspace}

\newcommand{\diff}{\mathrm{d}}

\newcommand{\GaussianAmp}{A}
\newcommand{\GaussianMean}{\mu}
\newcommand{\GaussianStd}{\sigma}

\newcommand{\NRMSE}{NRMSE\xspace}
\newcommand{\NMAE}{NMAE\xspace}

\newcommand{\EnergyNet}{MCLNet\xspace}
\newcommand{\TrajectoryNet}{TrajectoryNet\xspace}

\newcommand{\Pearson}{r}
\newcommand{\Spearman}{r}

\usepackage{color}
\usepackage{ifthen}
\usepackage{float}
\usepackage{alltt}
\usepackage{amsmath}
\usepackage{amsthm}
\usepackage{newlfont}
\usepackage{floatflt}
\usepackage{wrapfig}
\usepackage{fixltx2e}
\usepackage{subfig}
\usepackage{multirow}
\usepackage{booktabs}
\usepackage{cleveref}
\usepackage{mathtools, cuted}
\usepackage{CJKutf8}
\usepackage[ruled]{algorithm2e}
\usepackage{diagbox}
\usepackage{algpseudocode}
\usepackage[toc,page,titletoc]{appendix}
\crefname{supp}{Supplement}{Supplements}
\usepackage{enumitem}

\definecolor{lightgreen}{rgb}{0.56, 0.93, 0.56}
\definecolor{moonstoneblue}{rgb}{0.45, 0.66, 0.76}
\definecolor{randColor}{rgb}{0.9, 0.51, 0.1}

\SetAlFnt{\small}
\SetAlCapFnt{\small}
\SetAlCapNameFnt{\small}
\SetAlCapHSkip{0pt}
\IncMargin{-\parindent}
\newcommand{\Caption}[2]{\caption[#1]{{\em #1} #2}}
\let\oldcaption\caption
\renewcommand{\caption}[2][]{\oldcaption[#1]{{\em #1} #2}}

\definecolor{figred}{rgb}{1,0,0}
\definecolor{figgreen}{rgb}{0,0.6,0}
\definecolor{figblue}{rgb}{0,0,1}
\definecolor{figpink}{rgb}{1,0.63,0.63}

\newcommand{\pseudocode}{Algorithm}
\floatstyle{plain}
\newfloat{algorithm}{tbhp}{lop}
\floatname{algorithm}{\pseudocode}

\newcommand{\filename}[1]{\url{#1}}
\newcommand{\foldername}[1]{\url{#1}}

\let\oldparagraph\paragraph
\iffalse

\renewcommand{\paragraph}[1]{\oldparagraph{\textbf{#1}.}}
\else

\renewcommand{\paragraph}[1]{\oldparagraph{{#1}.}}
\fi

\ifdefined\email
\else
\newcommand{\email}[1]{\url{#1}}
\fi

\copyrightyear{2023} 
\acmYear{2023} 
\setcopyright{acmlicensed}
\acmConference[SIGGRAPH '23 Conference Proceedings]{Special Interest Group on Computer Graphics and Interactive Techniques Conference Conference Proceedings}{August 6--10, 2023}{Los Angeles, CA, USA}
\acmBooktitle{Special Interest Group on Computer Graphics and Interactive Techniques Conference Conference Proceedings (SIGGRAPH '23 Conference Proceedings), August 6--10, 2023, Los Angeles, CA, USA}
\acmPrice{15.00}
\acmDOI{10.1145/3588432.3591495}
\acmISBN{979-8-4007-0159-7/23/08}

\begin{document}
\title[Toward Optimized VR/AR Ergonomics]{Toward  Optimized  VR/AR  Ergonomics: \\  Modeling and Predicting User Neck \nobreak Muscle \nobreak Contraction}

%\author{Yunxiang Zhang, Kenneth Chen, Qi Sun}
%\affiliation{\institution{New York University, USA}\country{}}
%\email{{yunxiang.zhang,kennychen,qisun}@nyu.edu}\

\author{Yunxiang Zhang}
\affiliation{\institution{New York University}\country{USA}}
\email{yunxiang.zhang@nyu.edu}
\author{Kenneth Chen}
\affiliation{\institution{New York University}\country{USA}}
\email{kennychen@nyu.edu}
\author{Qi Sun}
\affiliation{\institution{New York University}\country{USA}}
\email{qisun@nyu.edu}
\begin{teaserfigure}
\centering
\subfloat[two head motion trajectories in VR]{
    \label{fig:teaser-user}
    \includegraphics[width=0.36\linewidth]{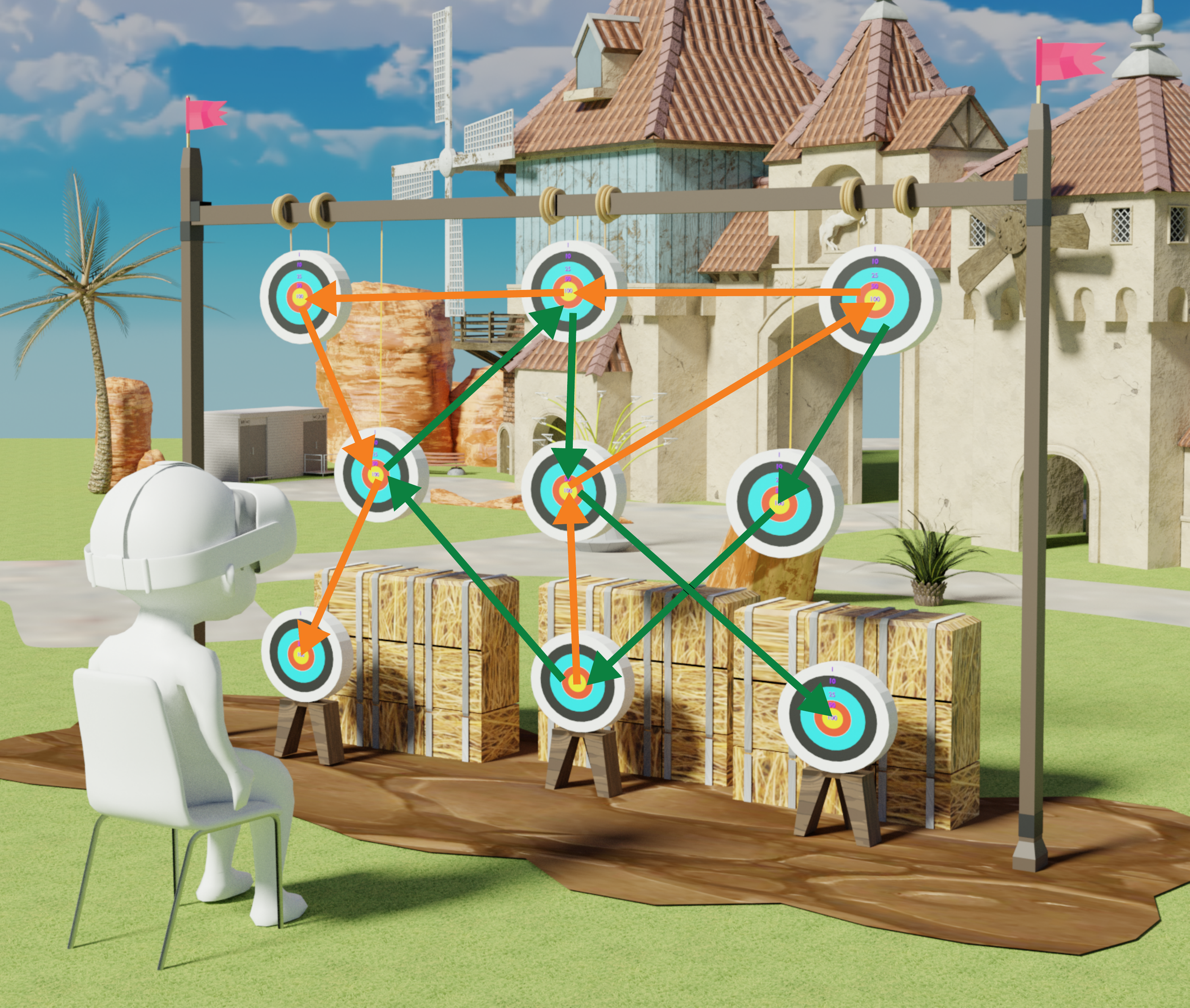}
  }
\subfloat[model-predicted cumulative neck muscle contraction and discomfort levels over time]{
    \label{fig:teaser-prediction}
    \includegraphics[width=0.62\linewidth]{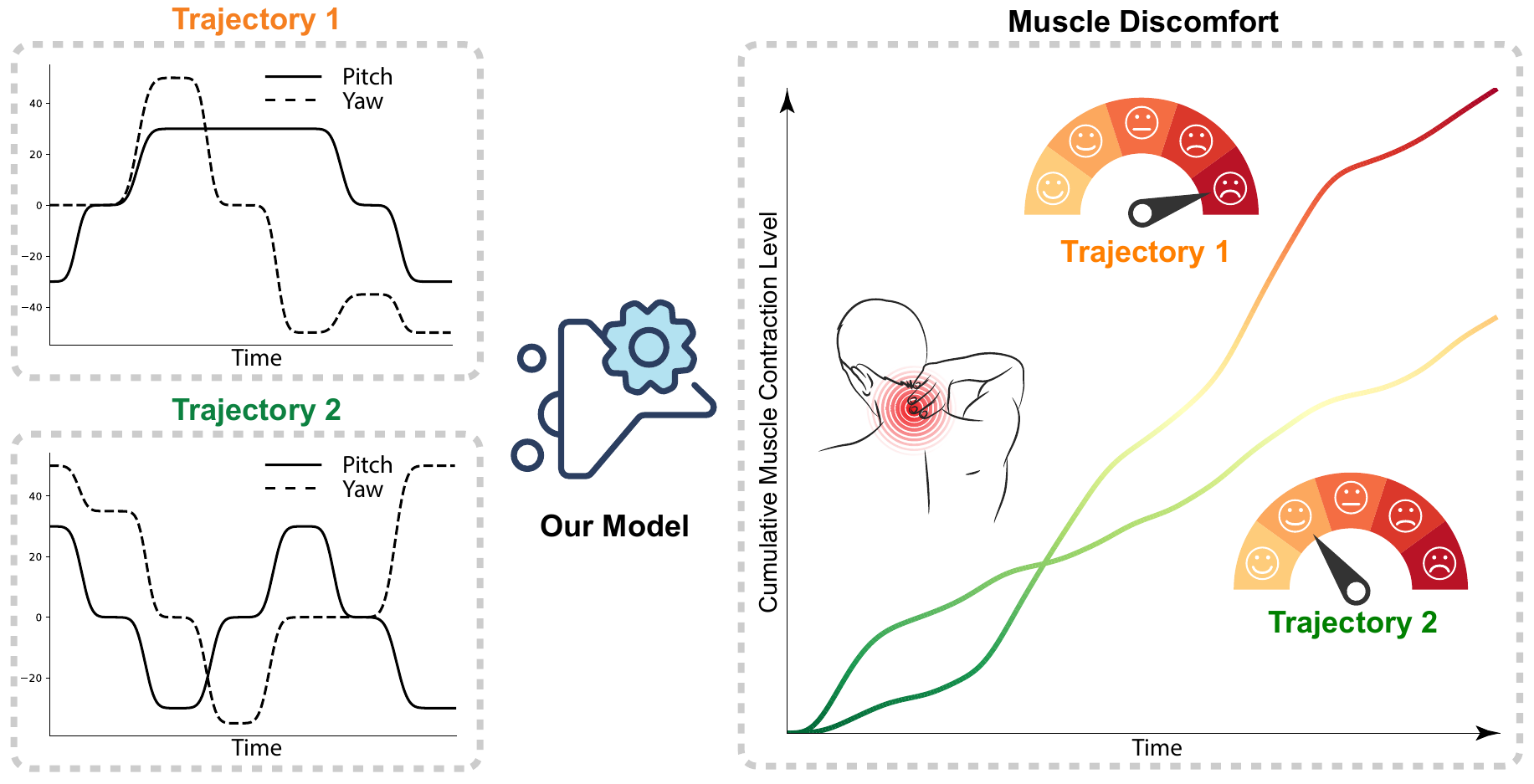}
  }
\caption[]{{Predicting the neck muscle contraction and discomfort levels of VR users.}
{\hyperref[fig:teaser-user]{(a)} A VR user chooses between two candidate head motion trajectories of seemingly similar muscular workload for a visual task. \hyperref[fig:teaser-prediction]{(b)} Our computational model predicts the user's potential neck muscle contraction level and thus perceived neck muscle discomfort before the movements happen. 3D asset credits to Mixall, Bizulka, RootMotion at Unity, and shockwavegamez01, joseVG at Sketchfab.}}
\label{fig:teaser}
\end{teaserfigure}

\begin{abstract}
Ergonomic efficiency is essential to the mass and prolonged adoption of VR/AR experiences. While VR/AR head-mounted displays unlock users' natural wide-range head movements during viewing, their neck muscle comfort is inevitably compromised by the added hardware weight. Unfortunately, little quantitative knowledge for understanding and addressing such an issue is available so far.

Leveraging electromyography devices, we measure, model, and predict VR users' neck muscle contraction levels (MCL) while they move their heads to interact with the virtual environment. Specifically, by learning from collected physiological data, we develop a bio-physically inspired computational model to predict neck MCL under diverse head kinematic states. Beyond quantifying the \nobreak cumulative MCL of completed head movements, our model can also predict potential MCL requirements with target head poses only. A series of objective evaluations and user studies demonstrate its prediction accuracy and generality, as well as its ability in reducing users' neck discomfort by optimizing the layout of visual targets. We hope this research will motivate new ergonomic-centered designs for VR/AR and interactive graphics applications. Source code is released at: \url{https://github.com/NYU-ICL/xr-ergonomics-neck-comfort}.
\end{abstract}




\maketitle
\section{Introduction}
\label{sec:intro}

VR/AR devices unlock natural viewing experiences via their uniquely wide-field displays. With head tracking, users can move their heads to shift attention and interact with peripheral content \cite{Bahill1975Saccade,monteiro2021hands}. However, their current head-mounted form factors incur non-trivial ``in vitro'' weight and shift the head's natural center of mass \cite{chen2021human}. The resulting changes in neck muscle state and workload have been evidenced to cause discomfort and injuries \cite{marklin2022head,forde2011neck,chihara2018evaluation,penumudi2020effects,souchet2022narrative}. Despite emerging evidence and concerns over such ergonomic side effects, comprehensively assessing and optimizing VR users' muscular comfort is still in its infancy.\enlargethispage{12pt}

A major cause of ergonomic discomfort is muscle fatigue and stress \cite{lowe1996pain}, especially from external weights, e.g., HMDs \cite{knight2004neck,chihara2018evaluation}. Unlike optically trackable body movements, measuring muscular activities is remarkably difficult. Besides indirect sensing such as calorimetry \cite{holdy2004monitoring}, biometrics from electromyography (EMG) sensors reveal muscular status via its detected electric potential generated by muscle fibers. In fact, the EMG signals directly correlate to muscle contraction \cite{komi1976signal}. Therefore, extensive literature attempted to understand our muscular functionalities during daily tasks, with face/gaze \cite{manssuer2016role}, arm/hand \cite{zhang2022force}, and full-body \cite{brown2021unified}.

Recent attention has arisen to measure the influence of emerging usage of HMDs \cite{chen2021human}. For instance, Chihara et al. \shortcite{chihara2018evaluation} measured and associated the altered muscular contraction with ergonomic discomfort by studying various viewing and interaction postures. However, surprisingly, we still have little quantitative knowledge of the introduced ergonomic effects before deploying a VR/AR application. Computationally forecasting muscle contraction is the foundation toward the ultimate aim - systematically optimizing visual content for ergonomically enhanced VR/AR.

We present a biophysically-inspired model to predict VR users' neck muscular contraction and thus potential ergonomic discomforts over time. The model is applicable to both after (given a head trajectory) and before (given a target position) users' head movements. We first perform a physiological study in VR to obtain EMG-sensed biometrics from  characterized natural head movements. The analysis reveals muscle contraction's significant correlations with head poses and motion patterns. Developed upon the data, our biophysical model first predicts the instantaneous muscle contraction given a head pose and angular acceleration. Then, by approximating representative trajectories \cite{farshadmanesh2012relationships}, the model further extends to forecast potential discomfort given only the target location and before the head movement occurs.

Our objective measurements and user studies demonstrate the model's: 1) prediction accuracy and generalizability with both post-hoc estimation and pre-hoc prediction, 2) capability in optimizing visual target layouts to reduce user-perceived muscular discomfort.

We hope this research will motivate new ergonomic-centered designs for VR/AR. As a first step, our model serves as a quantitative metric for evaluating and optimizing immersive applications, e.g., button layout in AR assistive tools or target positions in VR gaming.

In summary, our main contributions include:
\begin{itemize}
    \item an EMG-sensed biometrical dataset of VR users' neck muscle activity, characterizing wide ranges of head movements,
    \item a biophysically formulated and learned model that predicts muscular contraction with head poses and movements,
    \item an extended metric that forecasts the viewing-induced muscular efforts and discomfort level given the target position,
    \item demonstrations of the model's effectiveness in enhancing users' muscle comfort via altering targets' spatial layouts.
\end{itemize}
\section{Related Work}
\label{sec:prior}

%%%%%%%%%%%%%%%%%%%%%%%%%%%%%%%%%%%%%%%%%%%%%%%%%%%%%%%%%%%%%

\subsection{Ergonomics in VR/AR Interaction}
\label{sec:prior-ergonomics}

Ergonomic efficiency is essential to the mass adoption of VR/AR. Despite extensive efforts on designing more lightweight HMDs, their current form factors still considerably alter users' behaviors. Consequently, muscular discomfort \cite{forde2011neck,penumudi2020effects,chihara2018evaluation,souchet2022narrative}, especially in the neck and shoulders \cite{kim2018head,marklin2022head}, may be induced. Prior research has studied the impacts of head-supported mass on neck muscle activities under various application scenarios to better design dedicated hardware devices \cite{thuresson2003neck,thuresson2005mechanical,rubine2022sternocleidomastoid,le2021exploring}, such as military helmets. Complementing advances in hardware designs, we focus on modeling and predicting muscle activities under the VR/AR settings, where visual stimuli are controllable and optimizable, to guide the design of virtual content for better ergonomic comfort. Our approach shares a similar mindset to \cite{ruiz2018fine,li2020exertion}, i.e., task-dependent optimization of virtual content.\enlargethispage{12pt}

%%%%%%%%%%%%%%%%%%%%%%%%%%%%%%%%%%%%%%%%%%%%%%%%%%%%%%%%%%%%%

\begin{figure*}[t]
\newcommand{\pilotFigHeight}{4.0cm}
\centering
  \subfloat[neck muscle anatomy]{
    \includegraphics[height=\pilotFigHeight]{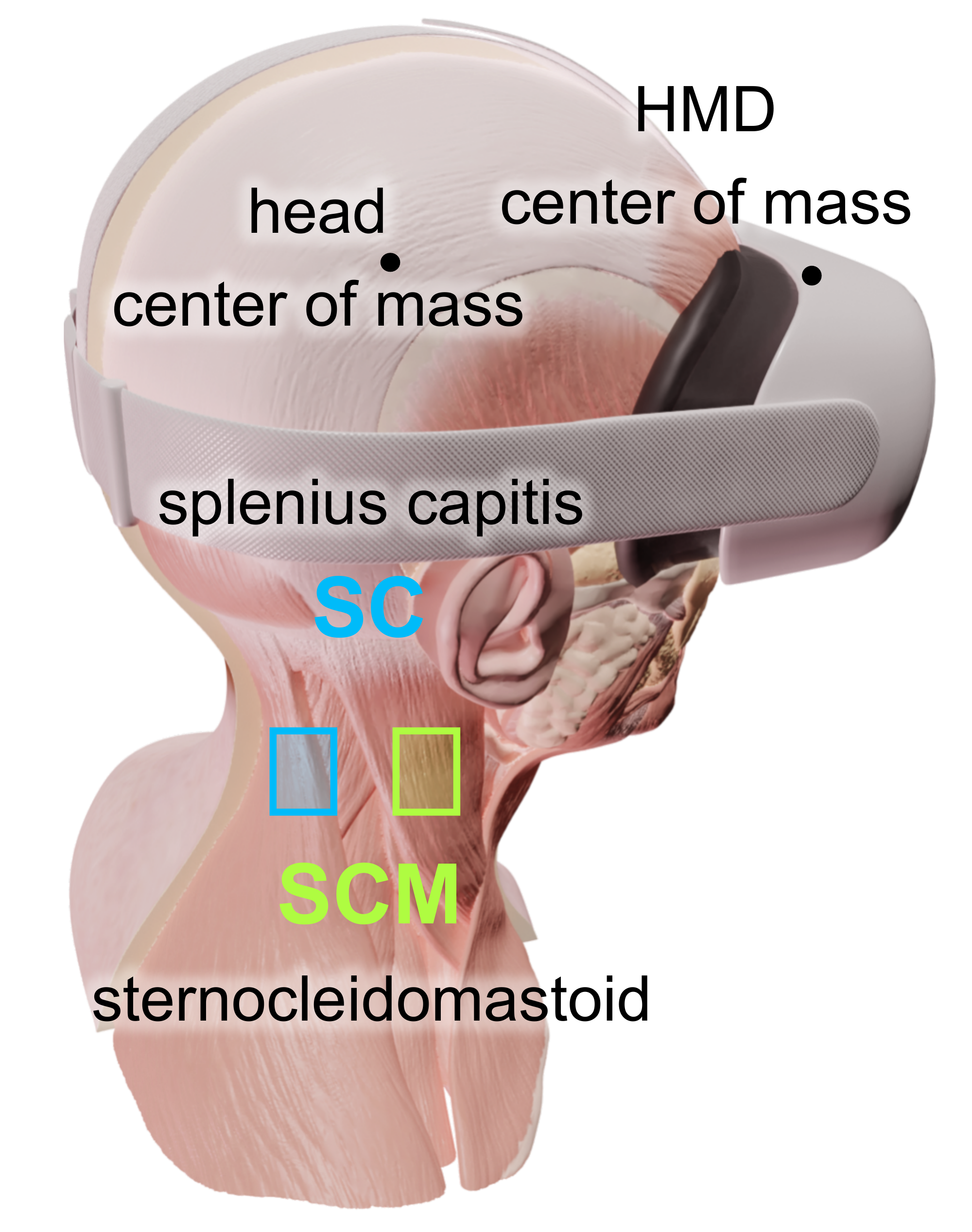}
    \label{fig:pilot:muscle}
  }%subfloat
  \subfloat[pilot study setup, EMG sensors, and stimuli]{
    \includegraphics[height=\pilotFigHeight]{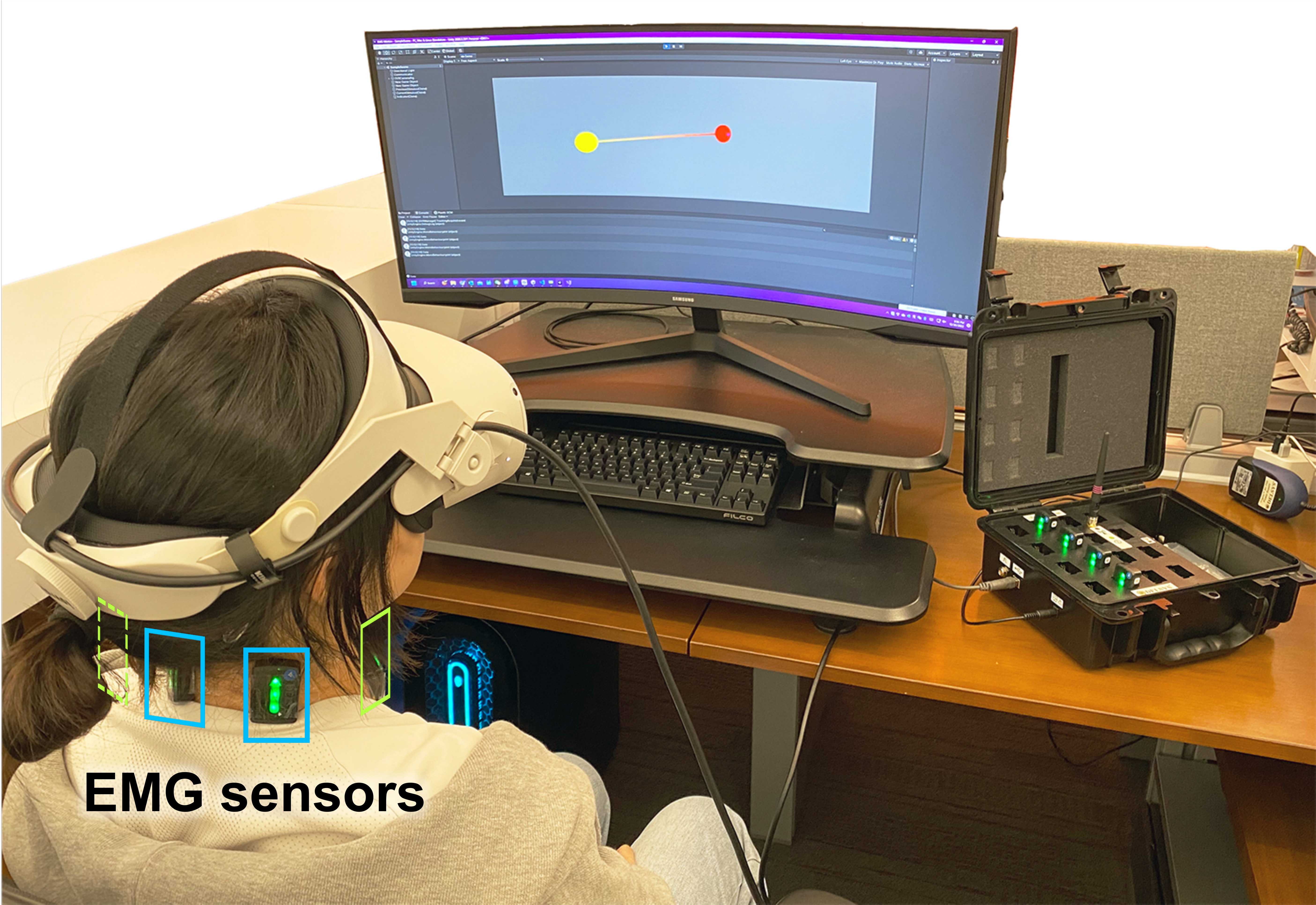}
    \label{fig:pilot:settingNstimuli}
  }%subfloat
  \subfloat[EMG data processing to obtain normalized \activationShort]{
    \includegraphics[height=\pilotFigHeight]{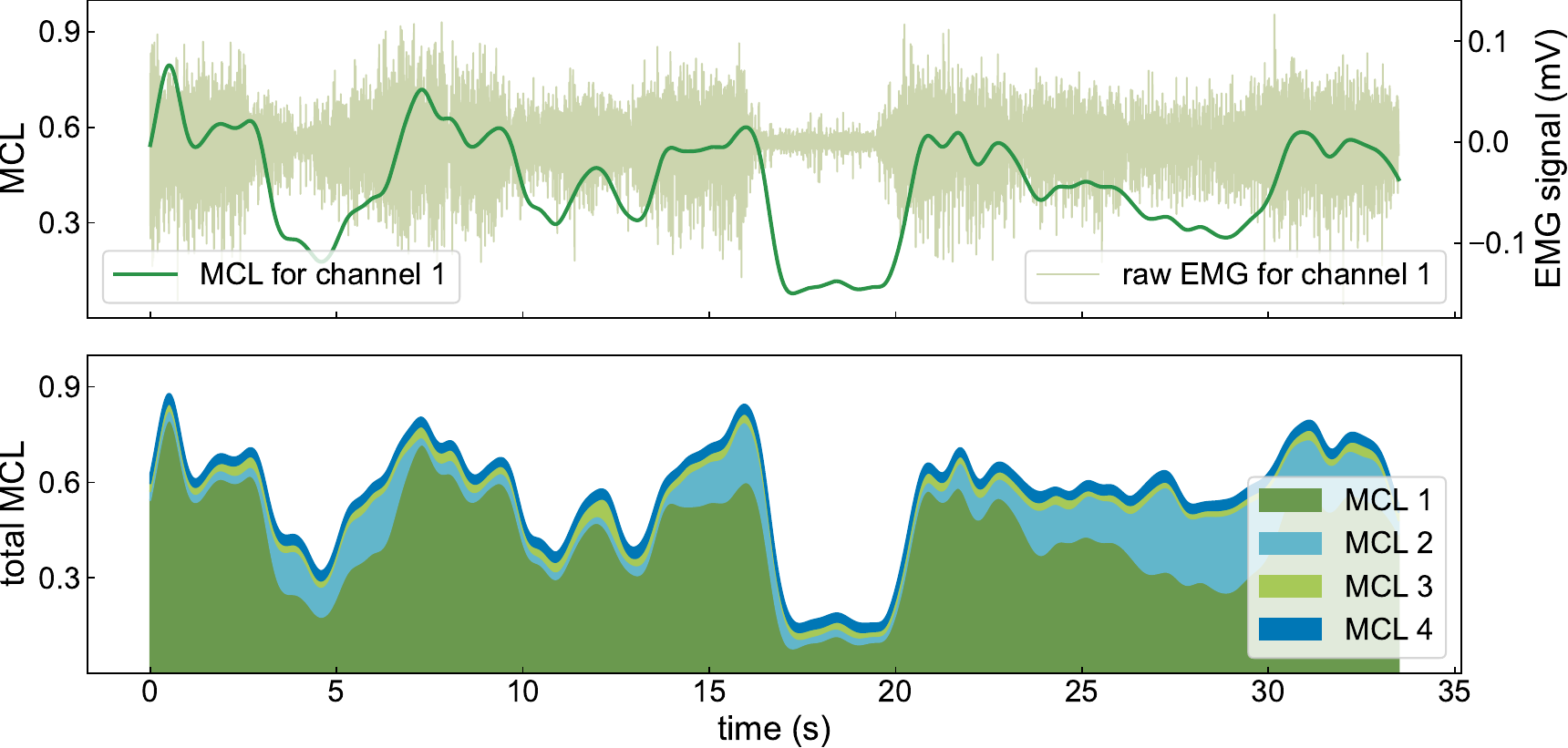}
    \label{fig:pilot:sample}
  } \\
  \subfloat[\activationShort during stationary viewing]{
    \includegraphics[height=4.2cm]{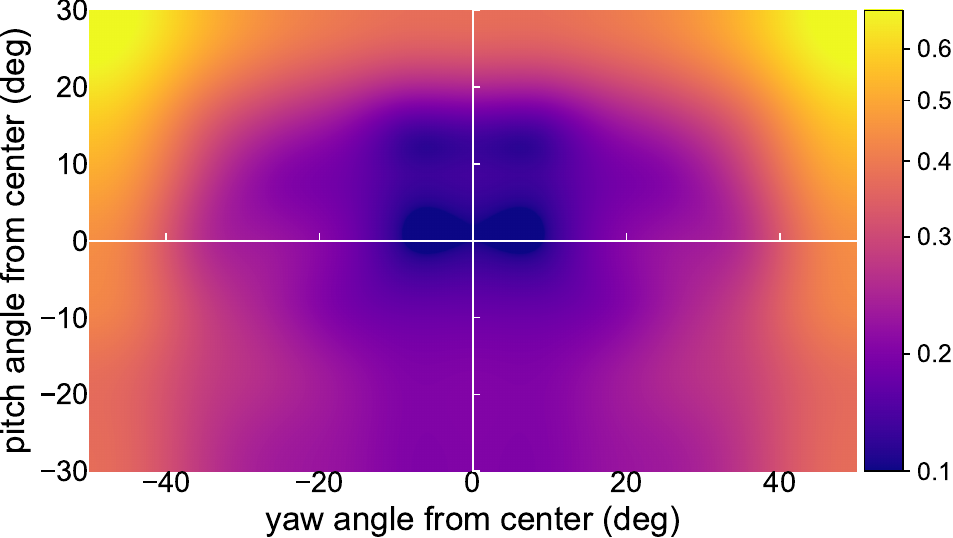}
    \label{fig:pilot:results}
  }%subfloat
  \subfloat[$\Delta$\activationShort during dynamic viewing (head movements)]{
    \includegraphics[height=4.2cm]{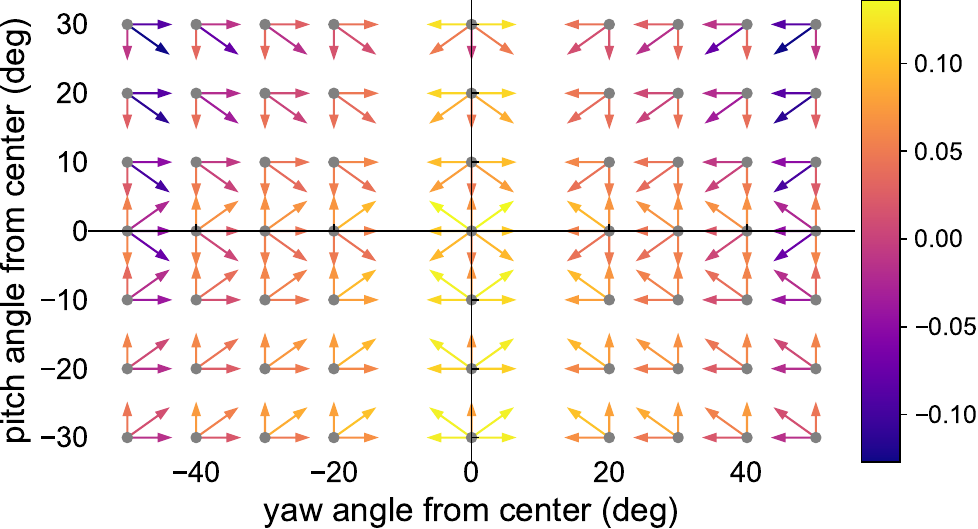}
    \label{fig:pilot:dynamic}
  }%subfloat
\caption[]{{Pilot study illustration and results.}
{
\hyperref[fig:pilot:muscle]{(a)} illustrates the major muscles controlling head movements, with highlighted EMG sensor attachment regions. 
\hyperref[fig:pilot:settingNstimuli]{(b)} shows our experimental setup with EMG sensors annotated. 
\hyperref[fig:pilot:sample]{(c)} The top row shows an example raw EMG sequence (light green curve and right Y-axis) and its corresponding normalized \activationShort (dark green curve and left Y-axis). The bottom row shows the total \activationShort integrated across all $4$ channels.
\hyperref[fig:pilot:results]{(d)} visualizes the user-aggregated \activationShort for stationary viewing.
\hyperref[fig:pilot:dynamic]{(e)} shows the movement-induced $\Delta$\activationShort for dynamic viewing. Each $\Delta$\activationShort, i.e., arrow, was computed by subtracting the \activationShort during stationary viewing at an anchor head pose from the average \activationShort during head movement from that anchor head pose (arrow tail) to a target head pose (arrow head). The arrow lengths were scaled to $13\%$ for easier visualization. 3D asset credits to joseVG and danielmclogan at Sketchfab.
}}%
\label{fig:pilot}
\end{figure*}

%%%%%%%%%%%%%%%%%%%%%%%%%%%%%%%%%%%%%%%%%%%%%%%%%%%%%%%%%%%%%

\subsection{Muscle Contraction during Movements}
\label{sec:prior-energy}

Muscle contraction level is a core biometrical indicator for studying ergonomics \cite{dugan2000muscle}. Unlike motions, which can be reliably tracked by cameras, quantifying muscle activities is notably more challenging. Existing work mainly exploits EMG to reveal muscle activities by detecting the electric signals propagating in neurally-activated muscles \cite{merletti2004electromyography,criswell2010cram}. In particular, elevated EMG readings indicate stronger muscle contraction and elevated discomfort over time \cite{chesler1997surface,dimitrova2003interpretation,cifrek2009surface,vigotsky2018interpreting}. Recently, researchers have explored machine learning techniques to infer muscle activities from spinal cord signals \cite{guo2018convolutional,gok2019prediction} and simulated human musculoskeletal animation data \cite{nakada2018deep}.

%%%%%%%%%%%%%%%%%%%%%%%%%%%%%%%%%%%%%%%%%%%%%%%%%%%%%%%%%%%%%

\subsection{Learning from Electromyography Signals}
\label{sec:prior-emg}

Learning EMG signals has emerged to understand human muscular behaviors with various applications \cite{ahsan2009emg,atzori2015ninapro,phinyomark2018emg}, including human-machine interfaces~\cite{moon2005wearable,atzori2016deep,xiong2021deep,karolus2022imprecise} and VR/AR~\cite{tsuboi2017proposal,hirota2018gesture,pai2019assessing,lou2019realistic}. EMG data has also been leveraged to enable various sensing tasks, such as body movement~\cite{jaramillo2020real,baldacchino2018simultaneous,zhao2020emg,javaid2021classification,du2017semi,wei2019surface}, hand~\cite{liu2021neuropose} and head~\cite{barniv2005using,sugiarto2021surface} tracking. The flexible and non-invasive design of the latest EMG creates new possibilities for understanding our behaviors invisible to cameras. For instance, estimating force \cite{becker2018touchsense,zhang2022force,gailey2017proof,bardizbanian2020estimating,martinez2020online,wu2021optimal}, prosthetic \cite{castellini2009surface,gulati2021toward} and gait \cite{papagiannis2019methodology,nazmi2019walking} control. We aim to achieve the inverse by predicting the EMG-measurable muscle status from human head movements.
\section{Neck Muscle Contraction Level during Head Movements in VR}
\label{sec:pilot}

%%%%%%%%%%%%%%%%%%%%%%%%%%%%%%%%%%%%%%%%%%%%%%%%%%%%%%%%%%%%%

\subsection{Neck Muscles Controlling Head Rotations}

We aim to model and predict neck muscle contraction level (\activationShort). Human neck is a highly flexible skeletal structure that allows the head to change its pitch, yaw, and roll angles. As shown in \Cref{fig:pilot}a, the major neck muscles include sternocleidomastoid (SCM) and splenius capitis (SC)~\cite{vasavada1998influence}. In particular, SCM/SC laterally rotate the head to the opposite/same side when acting unilaterally and flex/extend the head when acting bilaterally.\enlargethispage{12pt}

%%%%%%%%%%%%%%%%%%%%%%%%%%%%%%%%%%%%%%%%%%%%%%%%%%%%%%%%%%%%%

\subsection{Experiment Description}
\label{sec:pilot-experiment}

\paragraph{Participants and setup}
We recruited $8$ participants (ages $23-31$, $3$ female). 4 of them have prior experience with VR headsets before the study. All participants reported normal neck muscle conditions. For each participant, we attached $4$ Delsys Trigno wireless EMG sensors on the left/right SCM and SC muscles (see Figures~\ref{fig:pilot}a and \ref{fig:pilot}b). We originally tested with $6$ EMG sensors by also including the upper trapezius (UT) muscles. However, EMG sensors on the UT exhibited significantly weaker signals compared to the others and were thus excluded from the experiments. The EMG sensors detect the electric potential (in millivolts, mV) generated by users' muscle fibers at $2000$ Hz and stream the data to a PC with $<1$ms latency. During the study, every participant wore an Oculus Quest 2 head-mounted display (HMD), remained seated, and performed a target reaching task with visual stimuli. Their head poses were tracked by the HMD and streamed to the same PC for time synchronization with EMG. The HMD provides $1872 \times 1856$ resolution per eye at $90$ FPS, and $98^\circ/104^\circ$ vertical/horizontal field of view (FoV).\enlargethispage{12pt}

\paragraph{Stimuli and tasks}
As illustrated in \Cref{fig:pilot}b, the stimuli were sequentially displayed pairs of spheres ($3^\circ$ in the FoV), one colored yellow and the other red. The yellow sphere indicates an ``anchor'' head pose $\PilotTargetPosStart\triangleq(\PitchStart,\YawStart)$, where $\Pitch/\Yaw$ represent pitch and yaw angles. The red sphere indicates a ``target'' head pose that is $\PilotTargetPosDelta\triangleq(\PitchDelta,\YawDelta)$ away from $\PilotTargetPosStart$. While keeping their torso stationary, participants were instructed to rotate their heads from the yellow anchor to the red target. Once the user successfully fixated on the target for $2$ seconds consecutively (simulating stationary viewing), both spheres were shifted to continue with the next trial. A line connecting the two spheres was rendered to guide the user's head rotation to the next target, designed to eliminate potential errors for target searching. Please refer to our video for an example of the study process.

\paragraph{Conditions}
Across participants, the stimuli pairs appeared randomly with a pre-sampled anchor set $\PilotConditionSet\triangleq\{(\PilotTargetPosStart_i, \PilotTargetPosDelta_i), i=1,\dots\PilotConditionSize\}$. Given the comfortable range of human head rotations \cite{ruiz2018fine}, we sampled stimulus directions within a $60^\circ \times 100^\circ$ visual field. We tested and chose not to cover additional vertical range to ensure \nobreak participants' neck comfort and avoid sprains. We started with a preliminary test to determine the effective range of $\PilotTargetPosStart$. Our test with $3$ participants showed non-trivial EMG change for $|\YawStart| > 15^\circ$ and $|\PitchStart| > 5^\circ$. Therefore, using a $10^\circ$ step size, we sampled 63 anchor head poses for $\PitchStart\in\{\pm30^\circ,\pm20^\circ,\pm10^\circ,0^\circ\}$ and $\YawStart\in\{\pm50^\circ,\pm40^\circ,\pm30^\circ,\pm20^\circ,0^\circ\}$.
Here, $-/+$ indicates left/right or down/up from the head's forward-facing direction. For each of the $63$ anchor poses $\PilotTargetPosStart$, $8$ surrounding targets giving varying movement patterns $\PilotTargetPosDelta$ were studied. They were selected with $\PitchDelta\in\{\pm35^\circ,0\},\YawDelta\in\{\pm 25^\circ,0\}$. We discarded the conditions with target stimuli outside of the $60^\circ \times 100^\circ$ range. That is, we ensured $\PilotTargetPosStart+\PilotTargetPosDelta\in[-30^\circ,+30^\circ]\times[-50^\circ,+50^\circ]$.

\paragraph{Duration}
We split the study into $7$ sessions (about 5 minutes each) with enforced breaks in between to avoid posture drifting. Every session was monitored to ensure the subject remained stationary below the neck. The study, including hardware setup, pre-study instructions, warm-up session (30 discarded trials), and breaks, took about $2.5$ hours per participant. In total, we collected about 5 hours of time-synchronized motion-EMG paired data. 

\paragraph{Data processing and analysis}
\label{sec:model-data-processing}
We aim to model and optimize neck muscle contraction driving head movements. However, raw EMG signals cannot be directly used because they exhibit: 1) frequency-dependent sensory noise; 2) oscillations between negative and positive values; 3) left-right asymmetry due to sensor positioning error \cite{chihara2018evaluation,lehman1999importance}; 4) cross-user difference in scale for the same head movement. Therefore, similar to prior literature \cite{sommerich2000use,reaz2006techniques}, we performed a series of EMG signal processing, including detrending, bandpass filtering, and rectification, as well as inter-channel balancing, normalization, and integration. Please refer to Supplement A for details. At each time frame, our processing pipeline outputs a single normalized muscle contraction value, integrated across processed EMG signals from all $4$ channels. \Cref{fig:pilot}c illustrates the EMG-to-\activationShort transformation with an example sequence.

\subsection{Results}
\label{sec:pilot-data}

\paragraph{Stationary viewing ($|\PilotTargetPosDelta|=0$)} \Cref{fig:pilot}d shows the \activationShort when the head remains static.
The average normalized \activationShort was $.32\pm.12$. The head pose demanding the least \activationShort ($.17\pm.02$) was $\PilotTargetPosStart=(0,0)$, significantly lower than far-reaching poses. For instance, when the target was located at $\PilotTargetPos=(30^\circ,50^\circ)$, the \activationShort was higher at $.68\pm.09$. A repeated measures ANOVA showed that both  $\PitchStart$ and $\YawStart$ had a significant main effect on \activationShort ($F_{6,42} = 1.17, p < .001$ for $\PitchStart$, $F_{8,56} = 1.5, p < .001$ for $\YawStart$). A significant $\PitchStart \times \YawStart$ interaction effect was also observed ($F_{48,336} = 7.4, p < .001$). In particular, higher absolute values of yaw elevate the corresponding average \activationShort, from $.21\pm.09$ with $\YawStart=0^\circ$ to $.47\pm.15$ with $|\YawStart|=50^\circ$. A Mann-Kendall (M.K.) trend test showed a significant monotonic trend ($\tau=1.0, p<.05$). On the other hand, the effect from pitch was asymmetric and non-monotonic. The highest values of $\PitchStart$, $\PitchStart=+30^\circ/-30\circ$ induces \activationShort at $.49\pm.16$/$.27\pm.09$. An M.K. trend test did not show a significant monotonic trend of pitch angle's effect on \activationShort ($\tau=-.6, p=.07$).

\paragraph{Dynamic viewing ($|\PilotTargetPosDelta|>0$)} 
\Cref{fig:pilot}e visualizes the $\Delta$\activationShort during head movements, which was computed by subtracting the stationary \activationShort at an anchor from the average \activationShort during the movement. Introducing movements (i.e., non-zero $\PilotTargetPosDelta$) significantly elevated \activationShort up to $31.21\%$ across all studied $\PilotTargetPosStart$. In addition to $\PilotTargetPosStart$, movement pattern $\PilotTargetPosDelta$ jointly influences the observed \activationShort. A repeated measures ANOVA showed that $\PitchDelta \times \YawDelta$ has a significant main effect on \activationShort ($F_{3,21} = 23.44, p < .001$). For each time frame, we further extracted the angular acceleration in both directions $\AngularAcceleration\triangleq(\AngularAccelerationPitch,\AngularAccelerationYaw)$. Pearson correlation coefficients were computed to assess the relationship;
There were positive correlations between \activationShort and $|\AngularAccelerationPitch|$ ($\Pearson(12431)=.14,p<.001$), as well as \activationShort and $|\AngularAccelerationYaw|$ ($\Pearson(12431)=.038,p<.001$).
The elevation rate depends on individual $\PilotTargetPosStart$. For example, the rate in yaw direction with movement starting at $\PilotTargetPosStart=(30^\circ,50^\circ)$ was $88.9\%$ higher than $\PilotTargetPosStart=(0^\circ,0^\circ)$.

\subsection{Discussion}
\label{sec:pilot-discussion}

The analysis above leads us to several observations and motivations for learning a computational model. First, despite individual participants' variances in muscular strengths and sizes, the measured \activationShort shares consistent trends for each condition, both during static viewing and dynamic movements. Second, the head pose (yaw $\YawStart$ and pitch $\Pitch$) significantly influences \activationShort. In particular, despite the left-right symmetry with yaw, the pitch angle exhibits significantly asymmetric and non-monotonic effects on \activationShort. Lower pitch angles (i.e., heads facing downward) tend to reduce \activationShort. Third, in dynamic scenarios, increasing acceleration significantly elevates \activationShort. The elevation effect size depends on both the corresponding starting head pose and movement direction.

% Similar to prior art \cite{umberger2003model}, we hypothesize that the spatial-temporal muscle fiber contraction levels alter both head holding and moving torques, as reflected in the EMG-obtained \activationShort. These observations motivate us to develop a biophysically-inspired computational model by learning from our data.

\section{Method: Modeling and Predicting Neck Muscle Contraction Level}
\label{sec:method}

The analysis of EMG-motion paired data from our pilot study motivates us to establish a computational model correlating head movements with neck muscle contraction level (\activationShort). Note that while \activationShort can be measured using EMG sensors, they are 1) tedious and costly to deploy; 2) insufficient to \emph{forecast} \activationShort before a movement happens. Therefore, we first propose a bio-physically inspired \activationShort estimation model with open functions to characterize muscle-driven head motions in \Cref{sec:method-energy-model}. Using our collected data, we then fit the open functions with machine learning models to: 
\begin{enumerate}
    \item estimate the \activationShort associated with a completed head movement, i.e., \emph{after} a movement happens (\Cref{sec:method-model-prediction});
    \item predict the \activationShort for a potential movement using target directions only, i.e., \emph{before} a movement happens (\Cref{sec:method-trajectory-regression}).
\end{enumerate}

%%%%%%%%%%%%%%%%%%%%%%%%%%%%%%%%%%%%%%%%%%%%%%%%%%%%%%%%%%%%%

\subsection{Bio-Physically Inspired \activationShort Model}
\label{sec:method-energy-model}

Muscle-generated torque is proportional to \activationShort~\cite{watanabe2009normalized,clancy2011identification,paquin2018history}. Our neck muscles actively generate the required amount of torque to enable head rotation at varying speeds. Denoting this \textbf{a}ctive torque as $\ActiveTorque \in \EuclideanThree$, we establish a mapping $\TorqueToEnergy\left(\cdot\right)$ such that $\text{\activationShort}=\TorqueToEnergy\left(\ActiveTorque\right)$.

From our pilot study (\Cref{fig:pilot}d), maintaining the head stationary at various poses requires significantly different levels of \activationShort, and thus $\ActiveTorque$. To maintain stationary viewing, however, there must be another pose-dependent torque that counterbalances $\ActiveTorque$. We term this underlying torque as \textbf{p}assive torque $\PassiveTorque \in \EuclideanThree$. We hypothesize that $\PassiveTorque$ is induced by factors such as gravity and muscle relaxation. Notably, $\PassiveTorque$ exists during both stationary and dynamic viewing conditions, but only depends on head poses, i.e., $\PassiveTorque\triangleq\PassiveTorque\left(\Rotation\right)$. By contrast, we proactively generate $\ActiveTorque$ to perform stationary viewing (compensating $\PassiveTorque$) and change our head pose at wish for dynamic viewing. Therefore, through the moment of inertia $\Inertia \in \EuclideanOne$, $\ActiveTorque$ correlates both with the head pose $\Rotation$ and angular acceleration $\AngularAcceleration$:
\begin{align}
    \PassiveTorque\left(\Rotation\right) + \ActiveTorque\left(\Rotation,\AngularAcceleration\right) &= \Inertia \times \AngularAcceleration.
\label{eq:physics:torque}
\end{align}
This further transforms the mapping $\TorqueToEnergy\left(\cdot\right)$ between $\ActiveTorque$ and \activationShort:
\begin{align}
\begin{split}
\text{\activationShort} &= \left(\TorqueToEnergy\circ\ActiveTorque\right)\left(\Rotation,\AngularAcceleration\right) = \TorqueToEnergy\left(\Inertia\times\AngularAcceleration-\PassiveTorque\left(\Rotation\right)\right)
\\
&\triangleq\CombinedEnergyFunctional\left(\TorqueToEnergy,\PassiveTorque,\Inertia,\Rotation,\AngularAcceleration\right)\mapsto\EuclideanOnePos,
\label{eq:physics:energy}
\end{split}
\end{align}
where $\Inertia$, $\PassiveTorque\left(\cdot\right)$, and $\TorqueToEnergy\left(\cdot\right)$ are the unknowns that map $\Rotation$/$\AngularAcceleration$ to \activationShort.

%%%%%%%%%%%%%%%%%%%%%%%%%%%%%%%%%%%%%%%%%%%%%%%%%%%%%%%%%%%%%

\subsection{Estimating \activationShort with Complete Trajectories}
\label{sec:method-model-prediction}

\begin{figure}[t]
\centering
  \subfloat[illustration of {\EnergyNet}'s architectural design]{
    \includegraphics[width=0.96\linewidth]{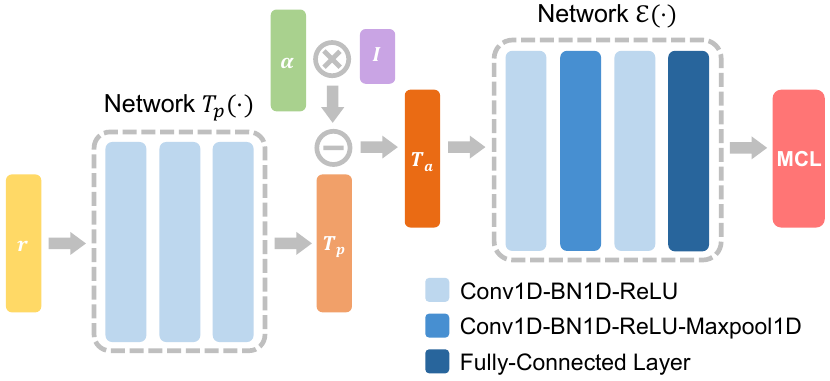}
    \label{fig:method:trajectory:nn}
  } \\ \vspace{-1em}
  \subfloat[model-predicted vs. hardware-measured neck \activationShort]{
    \includegraphics[width=0.96\linewidth]{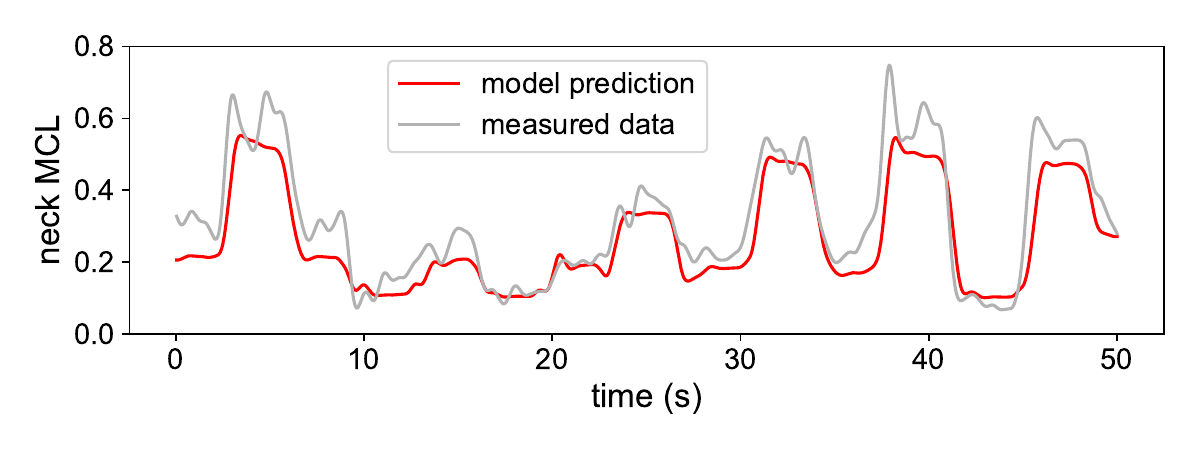}
    \label{fig:method:trajectory:prediction}
  }
\caption[]{\text{\EnergyNet} illustration and example \activationShort estimation results.
\hyperref[fig:method:trajectory:nn]{(a)} illustrates the architectural design of {\EnergyNet} for jointly learning $\Inertia$, $\PassiveTorque\left(\cdot\right)$, and $\TorqueToEnergy\left(\cdot\right)$ from head motion and \activationShort paired data; \hyperref[fig:method:trajectory:prediction]{(b)} shows the model-predicted vs. hardware-measured \activationShort for a sample sequence from the test set.}
\label{fig:method-model}
\end{figure}
 
The data from our pilot study provide a large set of time-synchronized head movement trajectories and \activationShort sequences:
\begin{align}
    \Rotation_\TimeStamp,\AngularAcceleration_\TimeStamp \mapsto \CombinedEnergyFunctional\left(\TorqueToEnergy,\PassiveTorque,\Inertia,\Rotation_\TimeStamp,\AngularAcceleration_\TimeStamp\right).
\label{eq:data}
\end{align}
Using these paired sequential data, we formulate an \activationShort regression problem and optimize 1D CNN models with $L_{2}$ loss to jointly approximate the unknowns $\Inertia$, $\PassiveTorque\left(\cdot\right)$, and $\TorqueToEnergy\left(\cdot\right)$. The complete model, named \textit{\EnergyNet}, is illustrated in \Cref{fig:method-model}a. 

\begin{figure*}[ht]
\centering
  \subfloat[Subject 1]{
    \includegraphics[width=0.49\linewidth]{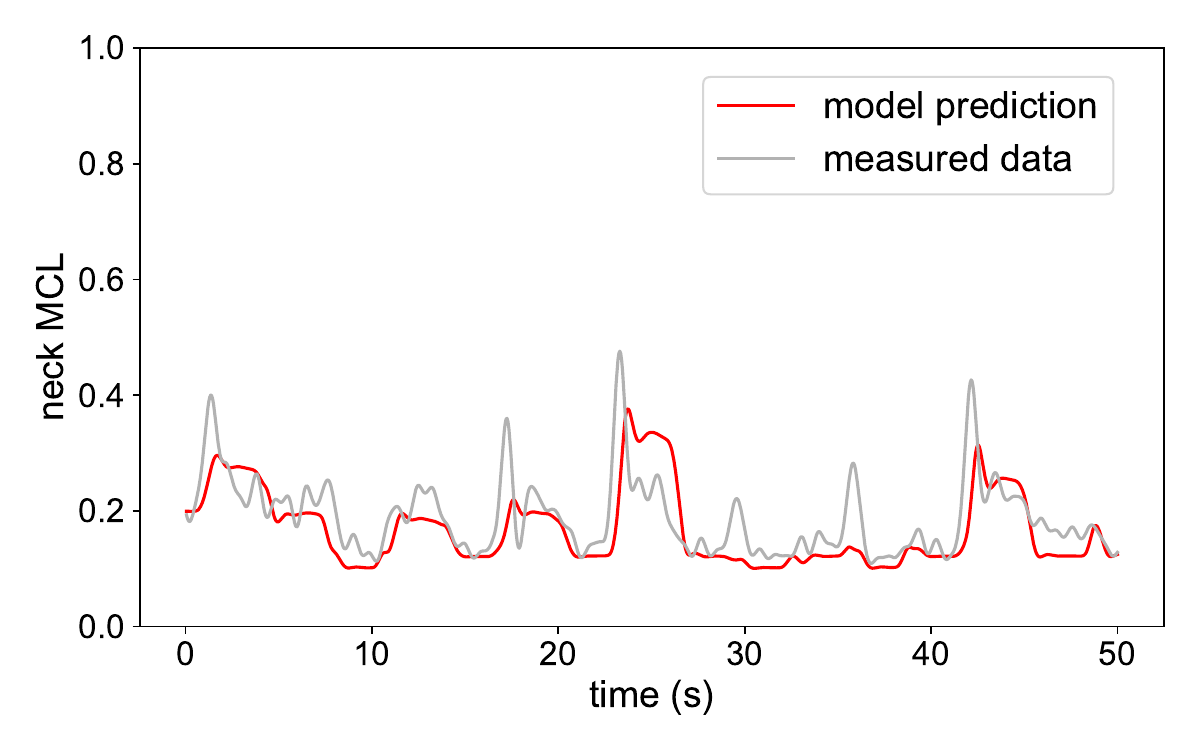}
    }
  \subfloat[Subject 2]{
    \includegraphics[width=0.49\linewidth]{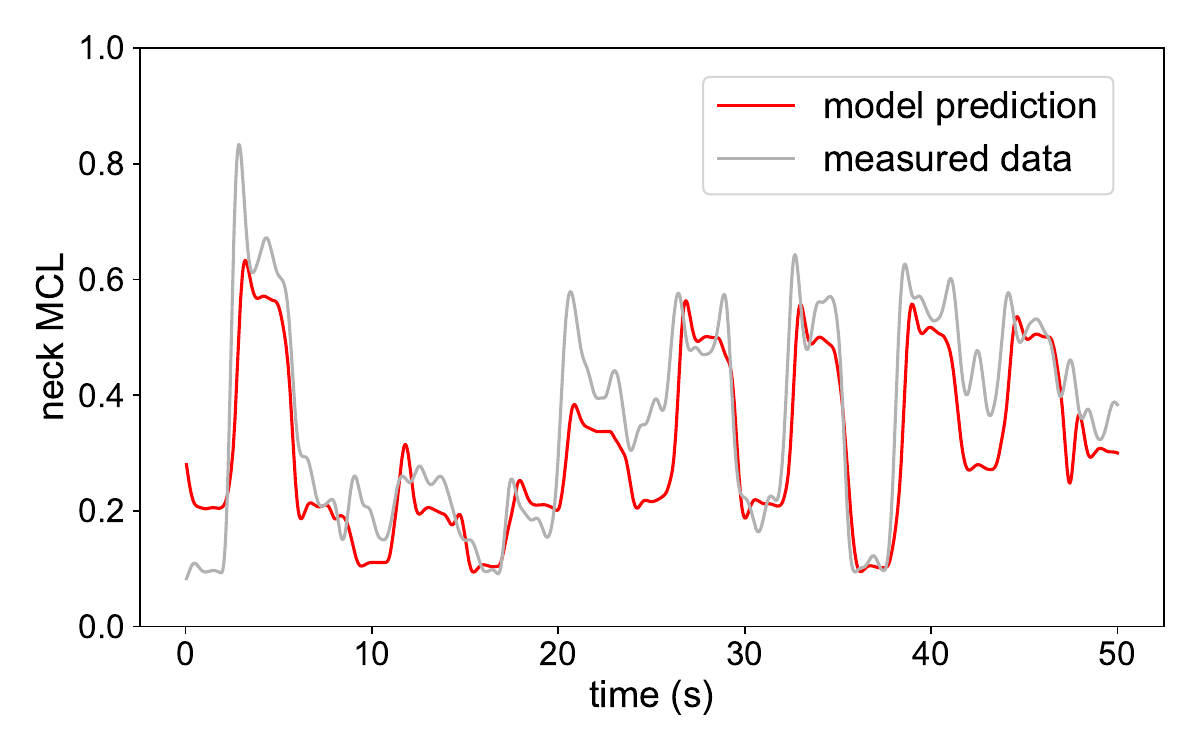}
    } \\
  \subfloat[Subject 3]{
    \includegraphics[width=0.49\linewidth]{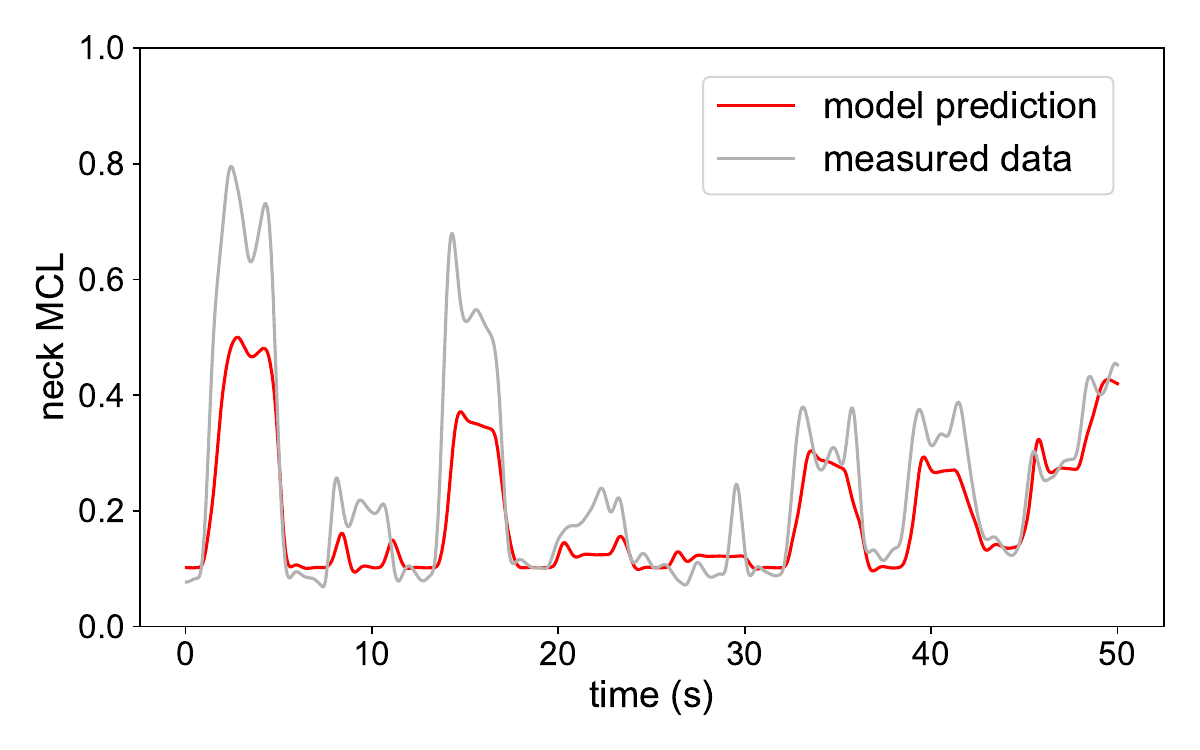}
    }
  \subfloat[Subject 4]{
    \includegraphics[width=0.49\linewidth]{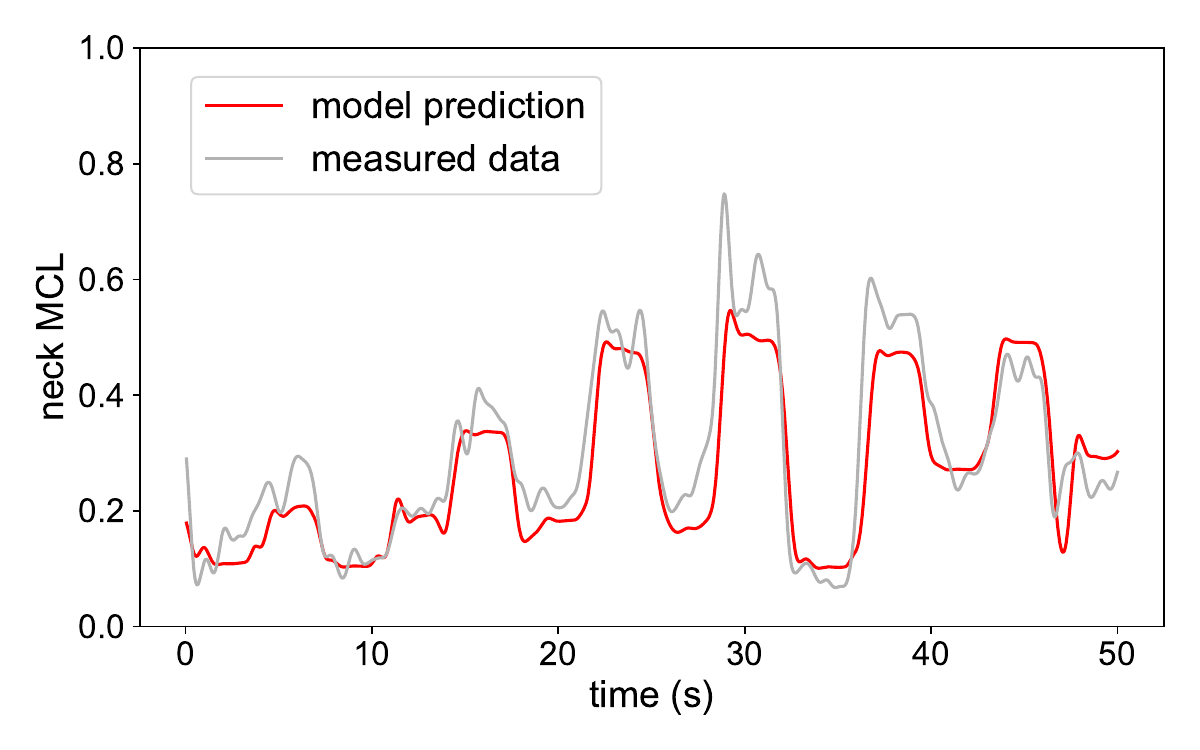}
    } \\
  \subfloat[Subject 5]{
    \includegraphics[width=0.49\linewidth]{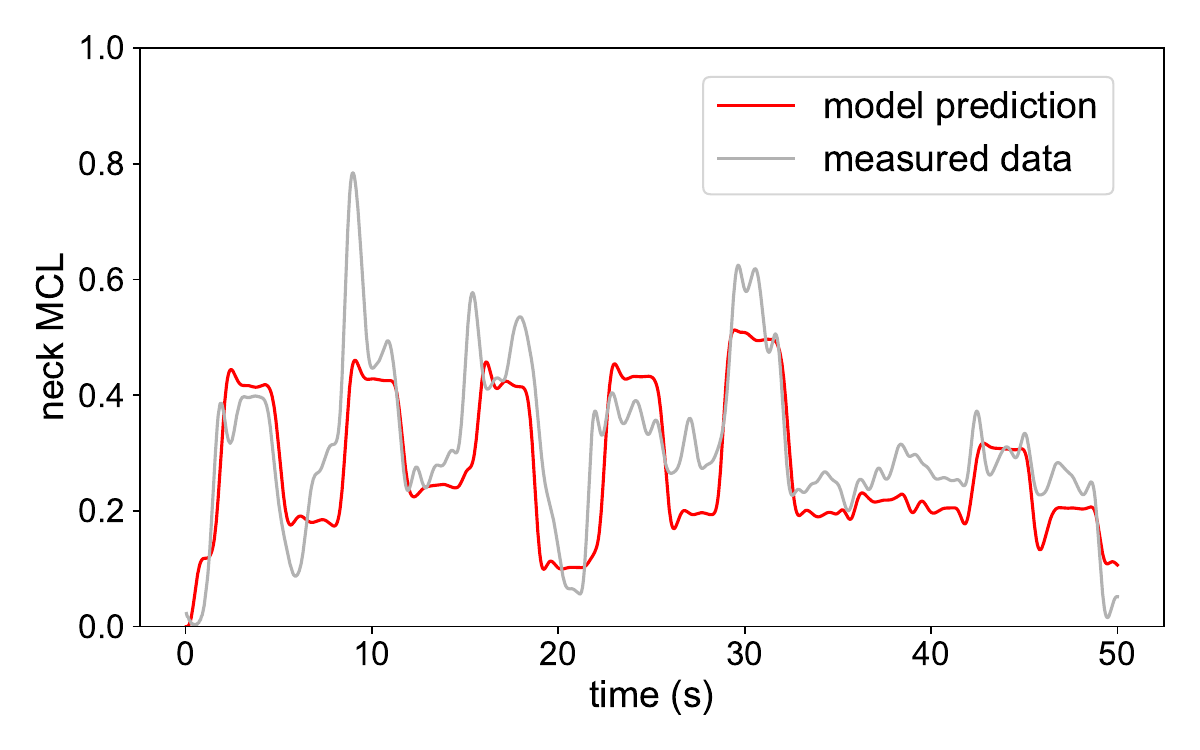}
    }
  \subfloat[Subject 6]{
    \includegraphics[width=0.49\linewidth]{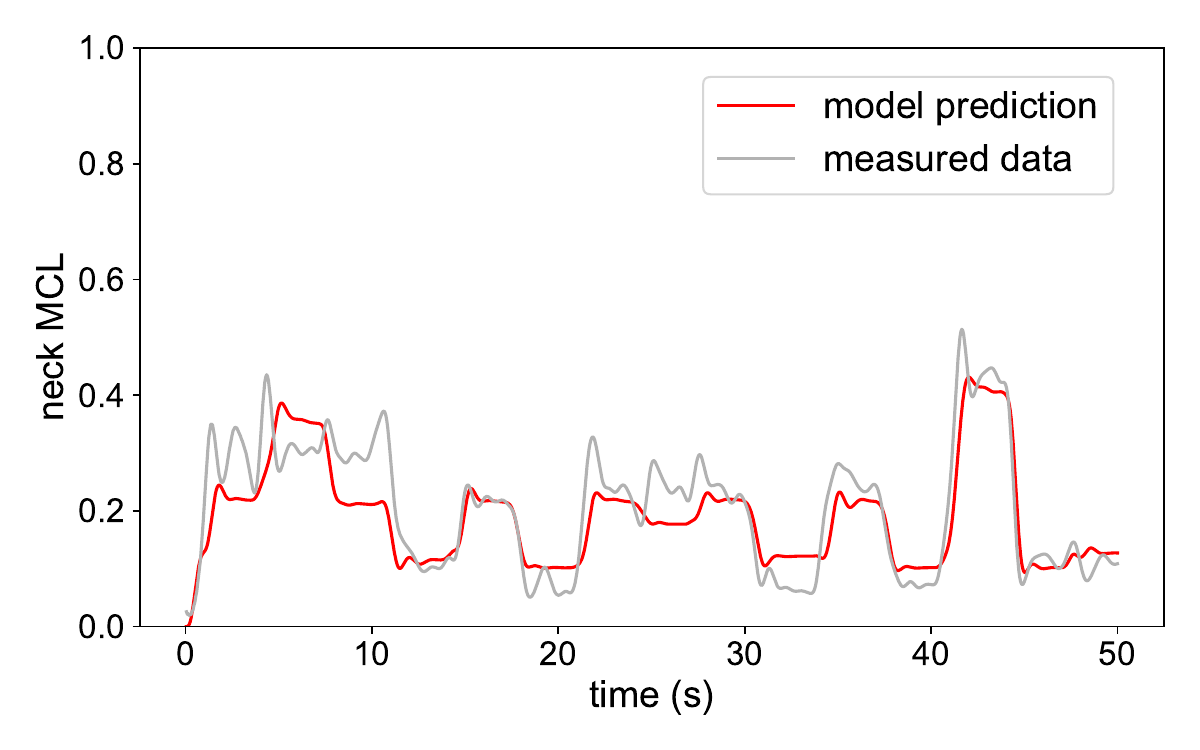}
    } \\
\caption[]{{Qualitative \activationShort prediction results.}{Model-predicted vs.
hardware-measured neck \activationShort for each of the 6 subjects who contributed evaluation data. Each sequence shown is randomly sampled from that particular subject's evaluation data.}}
\label{fig:results:curves}
\end{figure*}

Notably, a phenomenon called electro-mechanical delay exists between EMG signals and muscular motions. Depending on individuals and muscle areas, the delay can incur a temporal offset between the two modalities up to $100$ms~\cite{cavanagh1979electromechanical}. To accommodate this temporal inconsistency for robust prediction, our model takes in motion sequences with \TimeInterval=$400$ms window and predicts \activationShort for the central $200$ms interval, i.e., inputs cover additional $100$ms outputs from the beginning and end. Given a sequence of uniformly sampled head poses $\{\Rotation^{\TimeStamp}\}_{\TimeStamp=1}^{\TimeInterval}$, we first calculate the corresponding angular accelerations $\{\AngularAcceleration^{\TimeStamp}\}_{\TimeStamp=1}^{\TimeInterval}$ through finite difference, then execute our model at each $\TimeStamp$ to obtain the overlapping sequences of predicted \activationShort. \Cref{fig:method-model}b visualizes the prediction-measurement comparison for an example trajectory taken from the test set. Visualized results for each subject are shown in \Cref{fig:results:curves}.

So far, a core requirement of \EnergyNet is the prior knowledge of completed head movement trajectories. However, to benefit real-life applications such as UX design and cinematography, we shall reduce the potential discomfort before deploying to users. To this end, we further extend our model to \emph{forecast} \activationShort before a movement.

%%%%%%%%%%%%%%%%%%%%%%%%%%%%%%%%%%%%%%%%%%%%%%%%%%%%%%%%%%%%%

\subsection{Predicting \activationShort with Target Head Poses}
\label{sec:method-trajectory-regression}

\begin{figure}[thb]
\centering
\includegraphics[width=0.95\linewidth]{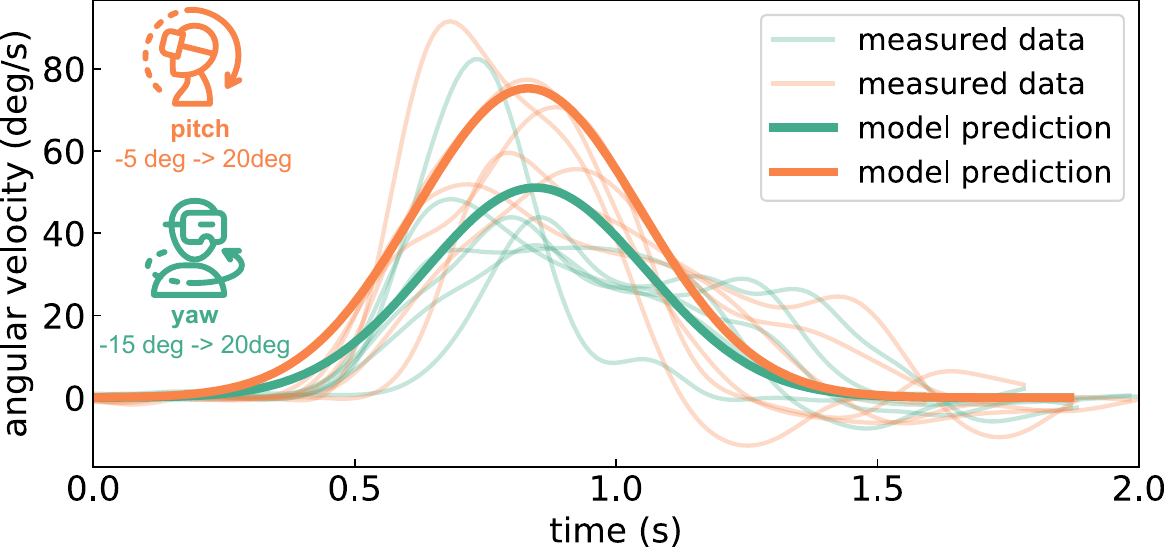}
\caption{{Predicting head motion trajectories with starting/ending head poses.} {Light/dark curves show tracked/predicted angular velocities. Orange/green curves show angular velocities in pitch/yaw directions.}}
\label{fig:method:combined}
\end{figure}

Given the starting and ending head poses $\{\RotationStart,\RotationEnd\}$ of a uni-directional head movement, the required \activationShort to travel between them is determined by the actual movement trajectory. However, as evidenced by our analysis in \Cref{sec:pilot-discussion}, $\{\RotationStart,\RotationEnd\}$ alone carry significant influence on the overall \activationShort. Therefore, using our collected data, we regress a representative motion trajectory for each pair of $\{\RotationStart,\RotationEnd\}$, to approximate the temporal patterns of angular velocity $\AngularVelocity_{\TimeStamp}$:
\begin{align}
\AngularVelocity_\TimeStamp(\RotationStart,\RotationEnd) \in \EuclideanThree, \text{\, s.t. \,}
\RotationStart + \int_{\TimeStamp=\TimeStampStart}^{\TimeStampEnd} \AngularVelocity_\TimeStamp(\RotationStart,\RotationEnd) \diff \TimeStamp = \RotationEnd.
\label{eq:velocity-integration}
\end{align}
Motivated by prior literature studying the main sequence effect of head movements \cite{zangemeister1981dynamics} and our observations of a single main peak in each velocity profile (\Cref{fig:method:combined}), we perform a unimodal Gaussian approximation for the angular velocity:
\begin{align}
\AngularVelocity_\TimeStamp^{i}(\RotationStart,\RotationEnd) \triangleq \GaussianAmp^{i}(\RotationStart,\RotationEnd) e^{-\frac{\left(\TimeStamp-\GaussianMean^{i}(\RotationStart,\RotationEnd)\right)^2}{2\left(\GaussianStd^{i}(\RotationStart,\RotationEnd)\right)^2}}, i \in \{\Pitch, \Yaw\}.
\label{eq:combined}
\end{align}
Using our collected data, we formulate a trajectory regression problem and optimize a Multi-Layer Perceptron (MLP) model, annotated as \textit{\TrajectoryNet}, to predict $\{\GaussianAmp^{i},\GaussianMean^{i},\GaussianStd^{i}\}_{i \in \{\Pitch,\Yaw\}}$ given an arbitrary pair of head poses $\{\RotationStart,\RotationEnd\}$. Using predicted angular velocity curves, we can approximate the overall \activationShort:
\begin{align}
\CombinedEnergy(\RotationStart,\RotationEnd)=\int_{\TimeStamp=\TimeStampStart}^{\TimeStampEnd}\CombinedEnergyFunctional\left(\TorqueToEnergy,\PassiveTorque,\Inertia,\RotationStart+\int_{\RepTimeStamp=\TimeStampStart}^{\TimeStamp} \AngularVelocity_{\RepTimeStamp}(\RotationStart,\RotationEnd)\diff\RepTimeStamp,\AngularVelocityDeriv_\TimeStamp\right)\diff\TimeStamp.
\end{align}
\Cref{fig:method:combined} compares the velocity curves collected from our users with {\TrajectoryNet}'s predictions over an example pair of $\{\RotationStart,\RotationEnd\}$ taken from the test set. Full implementation details for both \EnergyNet and \TrajectoryNet can be found in Supplement B.

\section{Evaluation}
\label{sec:evaluation}

We present a series of objective measurements on our model's performance in predicting neck \activationShort, and a subjective psychophysical study to demonstrate how the prediction reflects neck muscle discomfort. We first evaluate {\EnergyNet}'s estimation accuracy with complete head motion trajectories in \Cref{sec:results:trajectory}, then extend to predict \activationShort \emph{before} head movements by incorporating {\TrajectoryNet} in \Cref{sec:results:startEnd}. Through a user study, we demonstrate our method's potential in forecasting and reducing users' neck workload for a more comfortable VR experience in \Cref{sec:results:study}.

%%%%%%%%%%%%%%%%%%%%%%%%%%%%%%%%%%%%%%%%%%%%%%%%%%%%%%%%%%%%%

\subsection{\activationShort Estimation: After Head Movements}
\label{sec:results:trajectory}

\paragraph{Experimental setup}
We leveraged our data from the conditions detailed in \Cref{sec:pilot-experiment} to train our model. Additionally, during the pilot study, we also collected two groups of conditions to establish an evaluation dataset with unseen conditions. It contains $4$ pitch and $4$ yaw angles, $\PitchStart\in\{\pm25^\circ,\pm5^\circ\},\YawStart\in\{\pm45^\circ,\pm15^\circ\}$, resulting in a total of $16$ anchor head poses. The same $8$ surrounding targets with travel angle $\PitchDelta\in\{\pm35^\circ,0^\circ\},\YawDelta\in\{\pm 25^\circ,0^\circ\}$ were introduced to each of them. Note that the evaluation conditions were designed to contain no overlap with the training set. Due to the extra long collection process and thus scheduling conflicts, $6$ ($3$ female) of the $8$ participants completed the evaluation condition session. We adopted their data for this experiment. Two quantitative metrics were considered: Normalized Root-Mean-Square Error (\NRMSE) and Normalized Mean Absolute Error (\NMAE). The metrics are applied to measure the error ratio between the model-predicted and hardware-measured (by the same method detailed in Supplement A) \activationShort; a lower error ratio indicates better model performance.

\begin{figure*}[ht]
\centering
\includegraphics[width=0.7\linewidth]{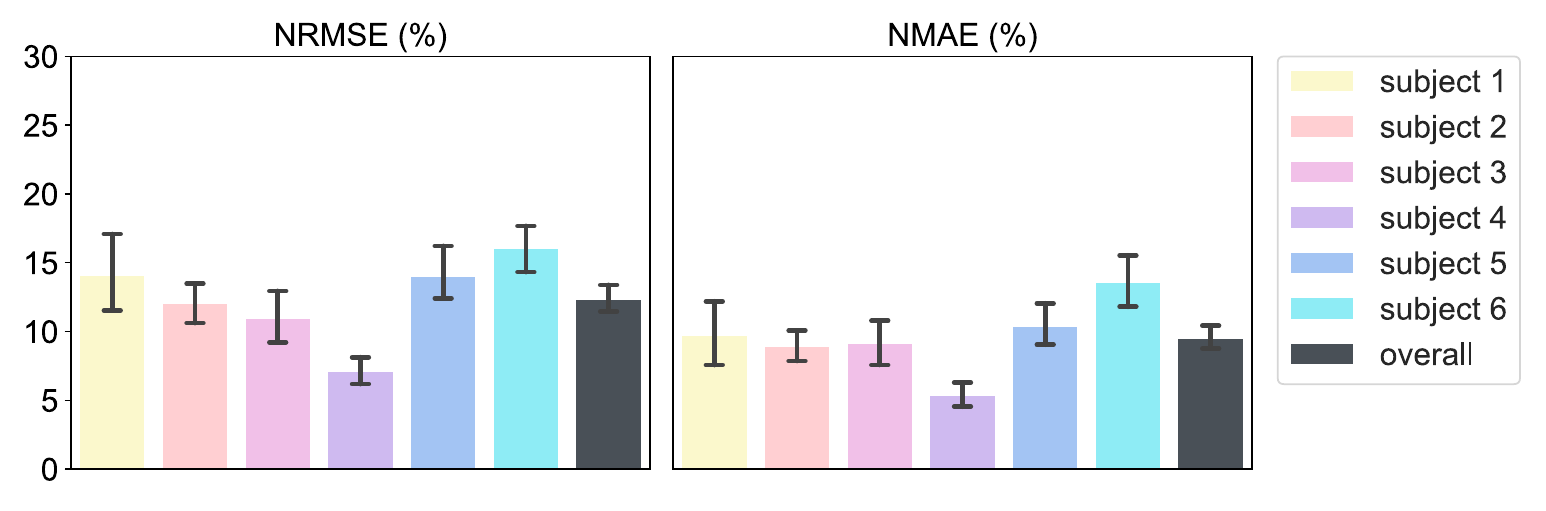}
\Caption{Performance of \EnergyNet for neck \activationShort estimation when complete head motion trajectories are known.}{}
\label{fig:results:motion}
\end{figure*}

\paragraph{Results and discussion}
\EnergyNet achieves an overall performance of $12.39\pm4.74\%$ \NRMSE and $9.54\pm4.14\%$ \NMAE across all 6 subjects and 16 anchor head poses. \Cref{fig:results:motion} summarizes its subject-wise performance. Beyond the average accuracy, we further measure the correlation. That is, whether the model can predict the elevation/reduction of \activationShort consistently with the hardware measurement. We leveraged Pearson's and Spearman's coefficients between the two conditions. The results indicate a significant correlation between model predictions and hardware measurements ($\Pearson(70066)=.62,p<.001$ and $\Spearman(70066)=.60,p<.001$). The analyzes above validate our method's effectiveness in estimating neck \activationShort when head motion trajectories were known beforehand.

%%%%%%%%%%%%%%%%%%%%%%%%%%%%%%%%%%%%%%%%%%%%%%%%%%%%%%%%%%%%%

\subsection{\activationShort Prediction: Before Head Movements}
\label{sec:results:startEnd}

\paragraph{Experimental setup}
For this experiment, we first used the head motion data from~\Cref{sec:results:trajectory} to optimize and validate \TrajectoryNet (\Cref{sec:method-trajectory-regression}), then combined it with \EnergyNet to predict \activationShort using target head poses only. Since our dataset consists of sequences of stationary head pose followed by pose-changing movements, we extracted the dynamic part to construct a dataset of starting/ending head poses $\{\RotationStart,\RotationEnd\}$ paired with HMD-tracked trajectories. The train-test data split from \Cref{sec:results:trajectory} was used for evaluation. 

\begin{figure*}[ht]
\centering
  \subfloat[performance of \TrajectoryNet for angular velocity prediction (pitch) with target head poses only]{
    \includegraphics[width=0.7\linewidth]{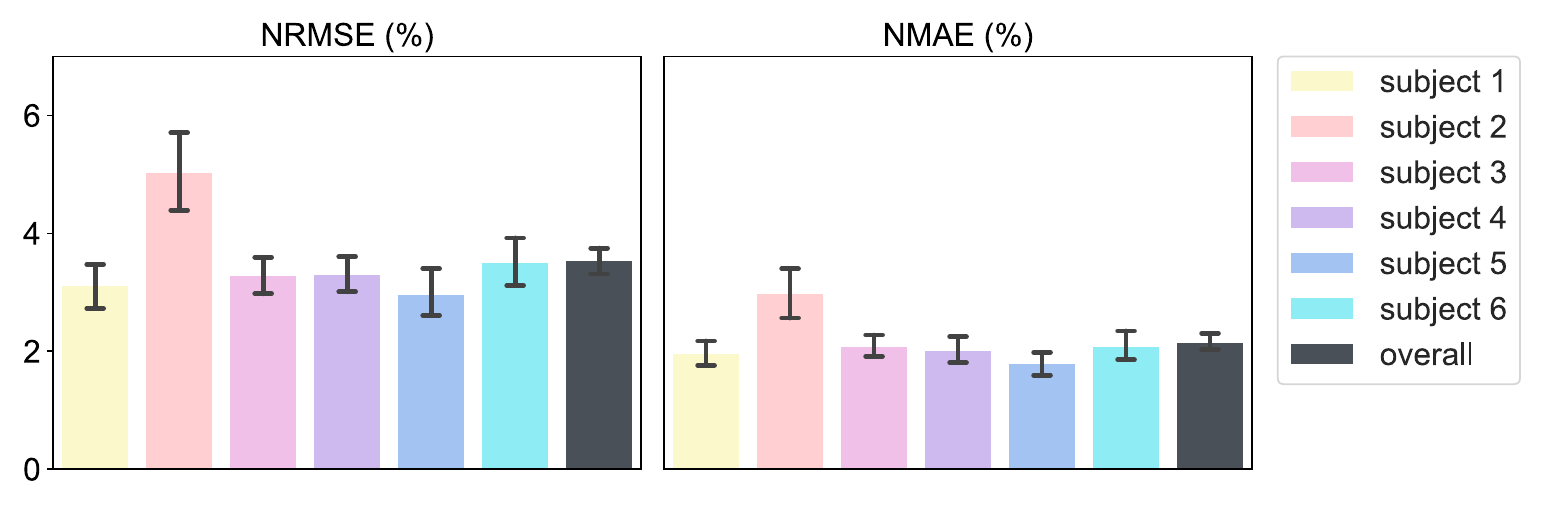}
    } \\ \vspace{-1em}
  \subfloat[performance of \TrajectoryNet for angular velocity prediction (yaw) with target head poses only]{
    \includegraphics[width=0.7\linewidth]{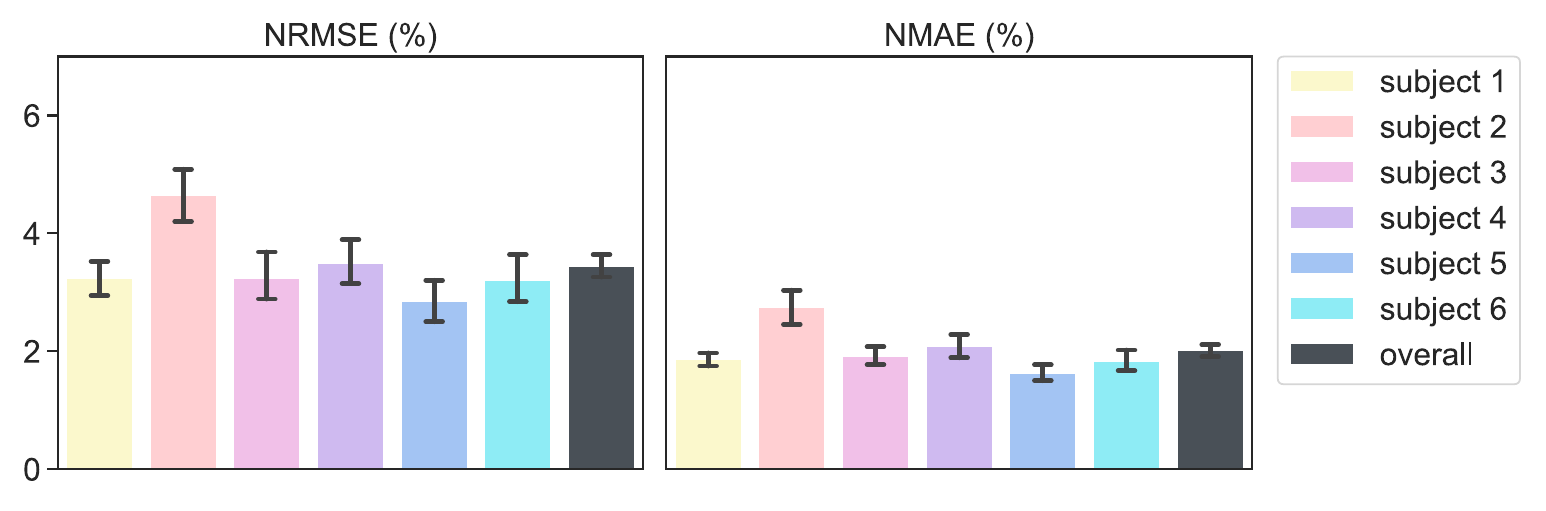}
    } \\\vspace{-1em}
  \subfloat[performance of \EnergyNet coupled with \TrajectoryNet for neck \activationShort prediction with target head poses only]{
    \includegraphics[width=0.7\linewidth]{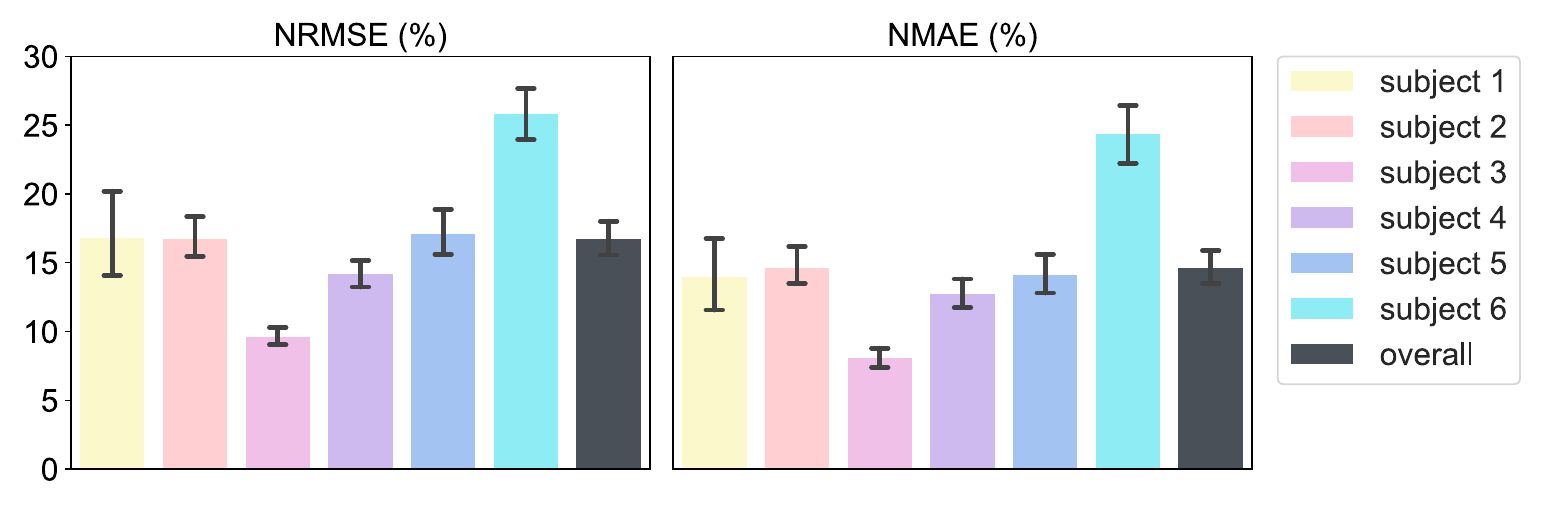}
    }
\caption[]{{Performance of our neck \activationShort prediction method (\EnergyNet$+$\TrajectoryNet) with target head poses only.}{Our method can reliably predict the potential neck \activationShort of a head movement before it takes place.}}
\label{fig:results:startEnd}
\end{figure*}

\paragraph{Results and discussion}
The performance of our \activationShort prediction framework, composed of trajectory regression and \activationShort estimation, is shown in \Cref{fig:results:startEnd}. \TrajectoryNet achieves an overall \NRMSE/\NMAE of $3.54\pm1.11\%$/$2.16\pm0.65\%$ in pitch velocity and $3.45\pm0.98\%$/$2.01\pm0.51\%$ in yaw. The overall \activationShort prediction performance is $16.76\pm6.05\%$ \NRMSE and $14.71\pm5.96\%$ \NMAE. Pearson's and Spearman's coefficients between the two conditions are $\Pearson(70066)=.59,p<.001$ and $\Spearman(70066)=.57,p<.001$, indicating a significant correlation. The results above demonstrate that our method can reliably predict the potential \activationShort with only the target head poses, before the actual movement occurs.

%%%%%%%%%%%%%%%%%%%%%%%%%%%%%%%%%%%%%%%%%%%%%%%%%%%%%%%%%%%%%

\subsection{Predicting and Reducing Neck Discomfort}
\label{sec:results:study}

\begin{figure*}[t]
\centering
  \subfloat[setting and stimuli]{
    \includegraphics[width=0.32\linewidth]{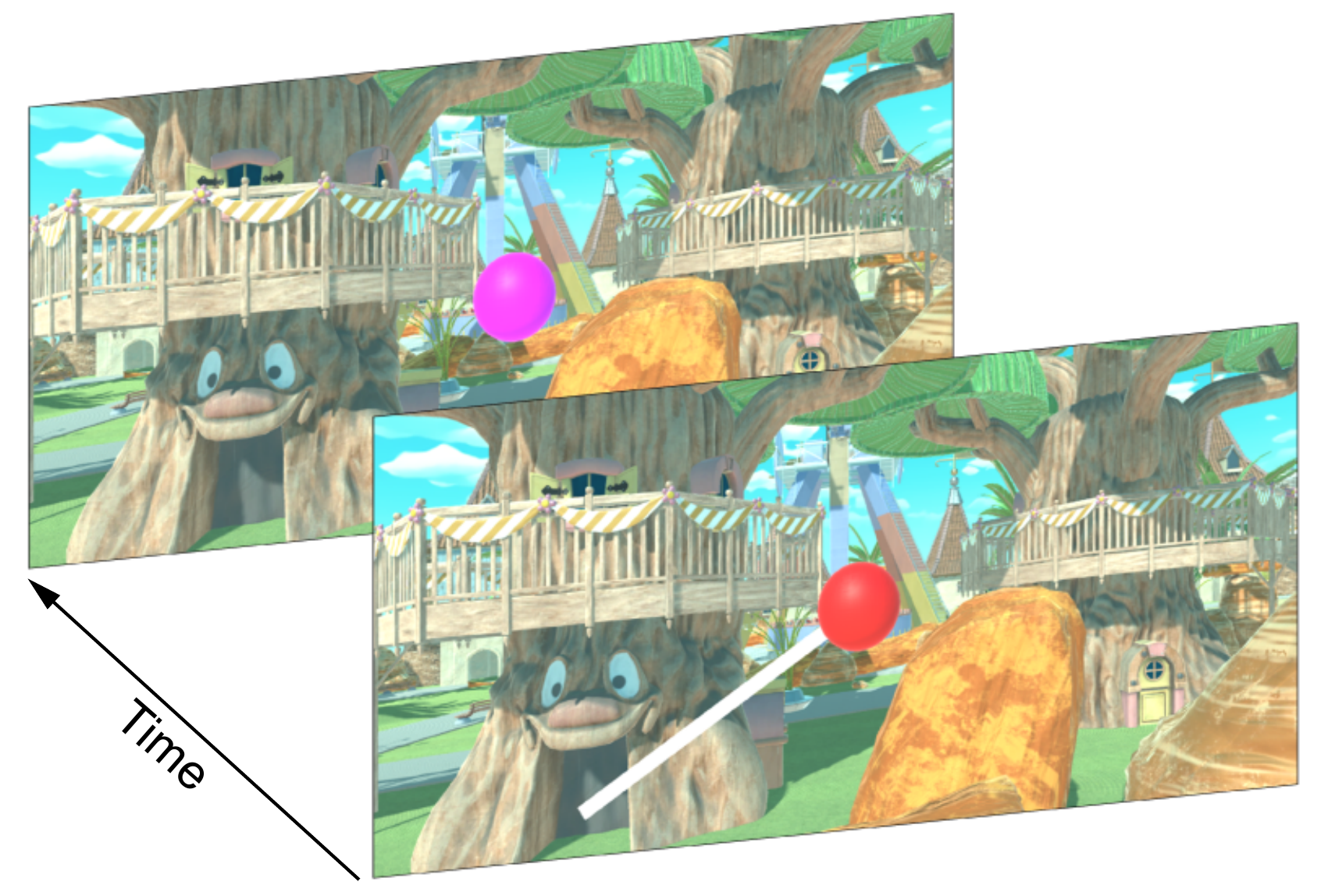}
    \label{fig:results:study:setting}
  }%subfloat
  \subfloat[visualization of conditions]{
    \includegraphics[width=0.32\linewidth]{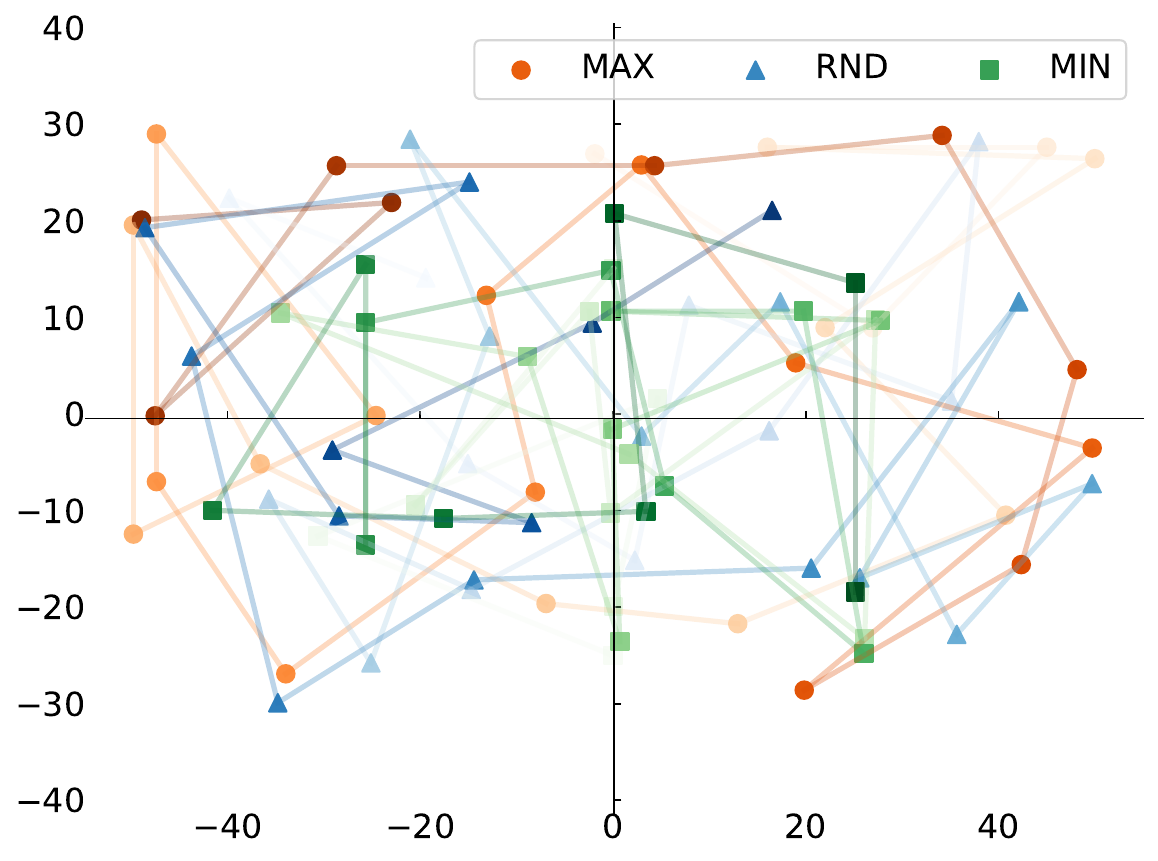}
    \label{fig:results:study:conditions}
  }
  \subfloat[aggregated voting results]{
    \includegraphics[width=0.32\linewidth]{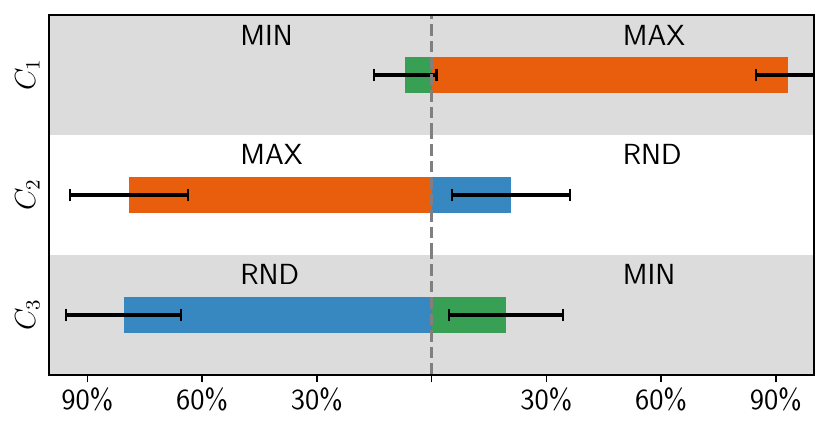}
    \vspace{-2em}
    \label{fig:results:study:results}
  }%subfloat
\caption[]{{Stimuli and results of our neck discomfort user study.}
{\hyperref[fig:results:study:setting]{(a)} shows the stimuli.
\hyperref[fig:results:study:conditions]{(b)} visualizes the visual targets' angular distributions of $3$ example conditions. Color gradients indicate the temporal order of appearance. All 3 conditions share the same total head rotations with full visual field coverage.
\hyperref[fig:results:study:results]{(c)} summarizes the voting distribution of the $3$ comparisons on which condition being more uncomfortable. Individual votes per condition are detailed in Table 1. 3D asset credits to Mixall at Unity.
}}
\label{fig:results:study}
\end{figure*}

\paragraph{Participants and setup}
We recruited $13$ participants (ages $20-35$, $6$ female). None of them were aware of the hypothesis, the research, or the number of conditions. One participant came in with a prior condition of neck injury and was excluded during warm-up sessions. There was no overlap between these participants and those from the pilot study in \Cref{sec:pilot}. We conducted the study using an Oculus Quest 2 HMD (\emph{without EMG}). During the study, participants observed stimuli through the HMD and were instructed to remain seated while keeping their torso stationary. The study took around $60$ minutes for each participant, including breaks between sessions.

\paragraph{Stimuli}
As shown in \Cref{fig:results:study}a, we developed and experimented with a 3D balloon-popping game with target acquisition. The stimuli were displayed as a sequence of red balloon targets rendered in an amusement park scene. As an indicator, the target color was changed to magenta upon a participant's fixation. A line pointing toward the next target was displayed to guide the user.

\paragraph{Tasks}
The task was designed as two-alternative forced choice to avoid bias from scaled rating, similar to prior works measuring muscular discomfort \cite{pinto2021motor,farid2018association}. During each session, the balloon targets were displayed, one at a time, following a pre-defined scan path with $31$ targets in different directions $\PilotScanPath=\{\PilotTargetPos_\TimeStamp, \TimeStamp = 0, \dots, 30, \PilotTargetPos_i \in \PilotConditionSet\}$, similar to the definition in \Cref{sec:pilot-experiment}. Participants were instructed to rotate their heads to fixate on the balloon until it disappeared. To trigger both dynamic and stationary head status, a 1-second fixation on each target was enforced before the next one appeared at $\PilotTargetPos_{\TimeStamp+1}$. Each session contained a pair of two sequentially tested $\PilotScanPath$ that were generated from $2$ out of $3$ different conditions, as detailed in the \emph{conditions} paragraph. After each session, the participants were instructed to use the keyboard to indicate ``which one of the two scan paths was more uncomfortable, tiring, or difficult for your neck?''.

\paragraph{Conditions}
Guided by our model, we designed $3$ conditions of progressively generated scan paths. They were created to ensure an identical total head rotations and similar spatial coverage but varied cumulative \expenditureShort. To ensure fair movements across conditions, we randomly pre-partitioned a fixed amount of total head rotations $900^\circ$ into $30$ steps. Then, at each step $\TimeStamp$, the next target pose $\PilotTargetPos_{\TimeStamp+1}$ was chosen from a set of candidate poses by maximizing the corresponding score function of the selected condition:
\begin{enumerate}[leftmargin=2cm]
\item[$\conditionMax$:]$\CoverageEnergy(\PilotTargetPos_{i\in[1,\TimeStamp+1]})+\CombinedEnergy(\PilotTargetPos_\TimeStamp,\PilotTargetPos_{\TimeStamp+1})$
\item[$\conditionRand$:] $\CoverageEnergy(\PilotTargetPos_{i\in[1,\TimeStamp+1]})$
\item[$\conditionMin$:]$\CoverageEnergy(\PilotTargetPos_{i\in[1,\TimeStamp+1]})-\CombinedEnergy(\PilotTargetPos_\TimeStamp,\PilotTargetPos_{\TimeStamp+1})$
\end{enumerate}
Here, $\CoverageEnergy$ is a term to ensure full visual field coverage for condition-wise fairness. On average, the ratio between the cumulative \activationShort of the three conditions $\conditionMax$/$\conditionRand$/$\conditionMin$ was $3.48$ vs. $1.95$ vs. $1.00$. Please refer to Supplement D for details on our condition generation algorithm and \Cref{fig:results:study}b for an example of each condition. The three conditions generate $3$ different pairs for 2AFC comparisons, namely $\conditionMaxVMin$: $\conditionMax$ vs. $\conditionMin$; $\conditionMaxVRand$: $\conditionRand$ vs. $\conditionMax$; $\conditionMinVRand$: $\conditionMin$ vs. $\conditionRand$. We repeated the random pre-partition of total head rotations and condition generation process to get 6 sets of $\conditionMax$/$\conditionRand$/$\conditionMin$ conditions for a total of 18 sessions. The appearance order of the 18 sessions was randomized and counter-balanced across participants, same for the 2 conditions within each session, to avoid bias.

\paragraph{Results}
Table 1 in Supplement E and \Cref{fig:results:study}c show individual votes and the summary for each comparison, respectively. By aggregating all sessions, $\conditionMax$/$\conditionRand$/$\conditionMin$ were $86.1\%$/$50.7\%$/$13.1\%$ voted as being more uncomfortable in the related comparisons. Among all comparisons, The difference was significantly higher than a random guess ($50\%$). By analyzing individual votes, a repeated measures ANOVA indicated that the condition had a statistically significant effect on the votes ($F_{2,22}=89.46, p<.001$). Post-hoc pairwise $t$-tests with Holm adjustments showed that the difference was significant among all $3$ comparisons ($p<.001$ for all conditions).

\paragraph{Discussion}
The analysis above shows the significant difference in participants' subjectively perceived discomfort levels among the three conditions. The participant-rated discomfort levels, $\conditionMax > \conditionRand > \conditionMin$, also matched our model's prediction ($\CombinedEnergy$). Note that the significant difference was not induced by head rotation angles which were ensured to be identical via our progressive trajectory generation. These results demonstrated our model's capability of predicting a user's neck discomfort with target head poses only, i.e., before the head movement takes place.
\section{Limitations and Future Work}
\label{sec:discussion}

This work considered the influence of yaw and pitch angles on \expenditureShort, but not the roll dimension due to the challenges of precisely manipulating it with visual stimuli and natural head movements. However, it may also contribute to  \expenditureShort \cite{keshner1989neck}. 
Introducing alternative tasks, such as full body movement \cite{imai2001interaction}, may enable controlling roll angles. We plan to investigate the options concerning their effects on noise and movement naturalness. Similar extensions include other muscle groups, such as shoulders during interaction \cite{chihara2018evaluation}.

In \Cref{sec:method-trajectory-regression}, we estimate the motion trajectory speed as a Gaussian representation given a starting and ending head pose. Despite the representativeness \cite{hage2019head}, the approximation may not fully contain the individuals' behavioral variances. We envision probabilistic modeling and learning \cite{ghahramani2015probabilistic} may further reveal the statistical variances across users.

%%%%%%%%%%%%%%%%%%%%%%%%%%%%%%%%%%%%%%%%%%%%%%%%%%%%%%%%%%%%%

\section{Conclusion}
\label{sec:conclusion}
Using EMG sensors, we present biometrically-measured data that reveals VR users' neck muscular contraction levels and thus potential discomfort.
By leveraging the data, we learn a computational model that quantitatively predicts the \expenditureShort, both after and before a head movement occurs. We hope the research to motivate new ergonomic and health-aware designs for VR/AR and interactive computer graphics, toward answering essential questions such as ``will VR/AR devices induce additional ergonomic burdens on users if they replace smartphones and monitors for everyday usage?'', ``how do we theoretically design more comfortable immersive displays and interfaces before they are deployed?''. To this end, our model may be applied to ergonomic-aware VR/AR interface optimization, immersive video editing, and beyond.
\begin{acks}
This project is partially supported by the National Science Foundation grants \#2225861 and \#2232817, and a DARPA PTG program.
\end{acks}

%%%%%%%%%%%%%%%%%%%%%%%%%%%%%%%%%%%%%%%%%%%%%%%%%%%%%%%%%%%%%

\clearpage
\bibliographystyle{ACM-Reference-Format}
\bibliography{paper.bib}

% \appendixpageoff
% \appendixtitleoff
% \renewcommand{\appendixtocname}{Supplementary material}
% \begin{appendices}
% \crefalias{section}{supp}
% \normalsize
% \begin{filecontents}{\jobname-support.tex}
% \clearpage
% \pagenumbering{roman}
% \input{sections/supplementary.tex}
% \end{filecontents}
% \include{\jobname-support}
% \end{appendices}

\end{document}